\documentclass[print, aps, prd, nofootinbib, superscriptaddress]{revtex4-2}
\usepackage[utf8]{inputenc}
\usepackage[T1]{fontenc}
\usepackage[english]{babel}
\usepackage{amssymb, amsmath, graphicx, epsfig, bm, appendix, braket}
\usepackage[caption=false]{subfig}
\usepackage{xcolor}
\usepackage{hyperref}
\definecolor{blue_refs}{rgb}{0., 0., 0.85}
\hypersetup{ 
colorlinks = true,
linkcolor = blue_refs,
citecolor = blue_refs,		
filecolor = blue_refs,		
urlcolor = blue_refs
}		
\DeclareMathOperator{\tr}{Tr}
\renewcommand{\d}{\mathrm{d}}
\newcommand{\e}{\mathrm{e}}
\newcommand{\sT}{{\scriptscriptstyle T}}
\newcommand{\myparallel}{{\mkern3mu\vphantom{\perp}\vrule depth 0pt\mkern2mu\vrule depth 0pt\mkern3mu}}

\begin{document}
\title{Azimuthal asymmetries in lepton and heavy-quark pair production in UPCs}

\author{Daniël Boer}
\email{d.boer@rug.nl}
\affiliation{Van Swinderen Institute for Particle Physics and Gravity, University of Groningen, Nijenborgh 3, 9747 AG Groningen, The Netherlands}

\author{Luca Maxia}
\email{l.maxia@rug.nl}
\affiliation{Van Swinderen Institute for Particle Physics and Gravity, University of Groningen, Nijenborgh 3, 9747 AG Groningen, The Netherlands}

\author{Cristian Pisano}
\email{cristian.pisano@unica.it}
\affiliation{Dipartimento di Fisica, Università di Cagliari, Cittadella Universitaria, I-09042 Monserrato (CA), Italy}
\affiliation{INFN, Sezione di Cagliari, Cittadella Universitaria, I-09042 Monserrato (CA), Italy}

\begin{abstract}
Azimuthal modulations in lepton and heavy-quark pair production in ultraperipheral collisions (UPCs) of highly charged ions are investigated. 
The modulations in the azimuthal angles of the sum and difference 
of the transverse momenta of the pair of particles in the final state, as well as of the transverse impact parameter, 
arise from the collisions of unpolarized and polarized photons.
A full description of the cross section in terms of Generalized Transverse Momentum Dependent parton distributions (GTMDs) for photons is given including a careful consideration of the Fourier transform to impact parameter space. In particular, this leads to a feed-in mechanism among harmonics of different orders, which in principle generates harmonics of all (even) orders. Wherever comparable, our analytical results for the azimuthal modulations agree with those presented in other papers on this topic. Compared to these other works, we separate effects that arise from the anisotropies of the GTMDs from those that do not and retain terms proportional to the mass of the produced particles, as they are relevant for muon, charm and bottom quark production. We show that the normalized differential cross section changes considerably with the produced particle mass, which should be discernible in UPCs at RHIC and LHC.  
For the numerical results we adopt several models for the photon GTMD correlator, and find that all of them are in fairly good agreement with each other and with UPC data from STAR. 
We also present results for various azimuthal modulations for RHIC kinematics, where we compare $e^+ e^-$ production with the production of heavier particles, and for LHC kinematics, focusing on $\mu^+ \mu^-$ production.
These results exhibit interesting mass-dependent features in the asymmetries that may help study the anisotropies arising from the underlying photon GTMD description.
\end{abstract}

\date{\today}
\maketitle

\section{Introduction}

Exclusive lepton-pair production in ultra-peripheral collisions (UPCs) of heavy ions was considered in a number of theoretical studies over the years \cite{Vidovic:1992ik,Hencken:1994my,Zha:2018ywo,Li:2019sin,Xiao:2020ddm,Wang:2021kxm,Mazurek:2021ahz,Shao:2022stc,Shao:2023zge,Shi:2024gex} with increased recent attention due to the RHIC data for this process \cite{STAR:2019wlg}. The process serves as a probe of the unpolarized and polarized photon distributions generated by a highly charged ion and allows to investigate in a rather clean way photon-photon interactions, similarly to the light-by-light scattering process first studied in UPCs by ATLAS \cite{ATLAS:2019azn}. 

Exclusive lepton-pair production in UPCs of heavy ions proceeds predominantly via the photon-photon fusion process $\gamma \gamma \to \ell^+ \ell^-$. The photon distribution of an ultra-relativistic charged ion (with atomic number $A$ and charge $Z$) is commonly described in the Equivalent Photon Approximation (EPA), first considered by Fermi, von Weizsäcker and Williams \cite{Fermi:1924tc,vonWeizsacker:1934nji,Williams:1934ad}. 
The distribution $f_{\gamma/A}(x)$ of photons with energy $\omega$ ($\hbar=1$) carrying a fraction $x = \omega/E$ of the momentum of a charged nucleon inside a nucleus is in this approximation given by the well-known result~\cite{Jackson:1998nia,Peskin:1995ev}\footnote{\label{ftn: 1} There is an extensive literature about the adequacy of the approximation and various modifications have been put forward, see e.g.\ Refs.~\cite{Brodsky:1971ud,Olsen:1979zb,Kniehl:1990iv,Frixione:1993yw}. For a recent discussion of this topic, see Ref.~\cite{Ma:2021lgv}.}:
\begin{align}
   f_{\gamma/A}(x) = \frac{Z^2 \alpha}{2\pi} \left(\frac{1+(1-x)^2}{x}\right) \ln\frac{E^2}{m^2}, 
\label{eq: photon distribution EPA}
\end{align}
where $m$ and $E$ denote the mass and energy of the charged nucleon. In the following $m$ will be denoted by $M_N$ to distinguish it from the mass $M_A$ of the nucleus. Supposing that the nucleus is moving (approximately) along the “$+$” direction, we have that the nucleon energy corresponds to $E = P_N^+/\sqrt{2}$. In the present case of UPCs of two heavy ions $E = \sqrt{s_{NN}}/2$, where $\sqrt{s_{NN}}$ is the center-of-mass (c.m.) energy per nucleon-nucleon collision. 
Note that the distribution of photons does not depend on the number of neutrons in this approximation, so in that sense $f_{\gamma/A}(x) = f_{\gamma/Z}(x)$.

The photon distribution of a nucleus in terms of its operator definition is given by 
\begin{align}
    f_{\gamma/A}(x) &= \frac{1}{x} \frac{1}{({\cal P}^+)^2} \int \frac{\d\lambda}{2\pi} \e^{i x  \lambda} \langle {\cal P} \vert \; F^{+\mu}(0)\, F^{+}{}_{\mu}(\lambda) \; \vert {\cal P} \rangle\,.
\label{eq: photon distribution definition}
\end{align}
Here ${\cal P}$ is the momentum of the nucleus, but $x$ is the fraction of momentum of a nucleon in the nucleus (where to good approximation $P_N={\cal P}/A$ and $x=A\, x_A$), such that without bound state effects $f_{\gamma/A}(x) = Z f_{\gamma/p}(x)+ (A-Z) f_{\gamma/n}(x)$, 
which for $f_{\gamma/n}(x)=0$ agrees with $f_{\gamma/A}(x) = Z f_{\gamma/p}(x) = f_{\gamma/Z}(x)$. 
The nucleus-nucleus cross section is given in terms of the photon-photon cross section
\begin{align}
\sigma_{AA} = \int \d x_1\, \d x_2 \, f_{\gamma/A}(x_1)\,  f_{\gamma/A}(x_2) \, \sigma_{\gamma \gamma}.
\label{eq: photon-photon cross section}
\end{align}
However, this expression implicitly involves integration over the impact parameter $b_\sT \equiv \vert \bm b_\sT\vert$, which is the distance between the colliding nuclei transverse to the beam axis. Hence one rather wants to describe $\d\sigma_{AA}/\d b_{\sT}$ in terms of the impact parameter dependent generalized photon distribution $f_{\gamma/A}(x,b_\sT)$, which is related to a Generalized Parton Distribution (GPD) with zero skewness, where only formally $f_{\gamma/A}(x) = \int \d^2 \bm b_{\sT}\, f_{\gamma/A}(x, b_\sT)$.
In UPCs, i.e.\ at large $b_\sT$, with Lorentz factor $\gamma = \omega/(x\, m)$, one has \cite{Jackson:1998nia, Vidovic:1992ik}:
\begin{align}
    f_{\gamma/A}(x,b_\sT) = \frac{Z^2 \alpha}{2\pi} \frac{m}{b_\sT} e^{-2 x m b_{\sT}}\, ,   
\label{eq: pointlike photon distribution}
\end{align}
but at smaller $b_\sT$ one will start to notice that the nucleus is not a point charge, such that one expects $f_{\gamma/A}(x,b_\sT)$ to depend on the number of nucleons (including neutrons) through the nuclear radius, and at some point (for more central collisions) the quarks and gluons inside the nucleus will play the dominant role.

The description of $\d\sigma_{AA}/\d b_\sT$ in terms of $f_{\gamma/A}(x,b_\sT)$, or rather the density $n(\omega,b_\sT)$ that is related to the photon distribution through $\omega\, n(\omega,b_\sT) = x\, f_{\gamma/A}(x,b_\sT)$, was considered in several studies~\cite{Vidovic:1992ik,Hencken:1994my,Zha:2018ywo}, but recently the further extension with transverse momentum dependence has received quite some attention~\cite{Li:2019sin,Xiao:2020ddm,Wang:2021kxm,Mazurek:2021ahz,Shao:2022stc,Shao:2023zge,Shi:2024gex}.
If one also measures the transverse momentum distribution of the lepton pair, the cross section becomes an expression in terms of the generalized photon distribution $f_{\gamma/A}(x,\bm{k}_\sT,\bm{b}_\sT)$ which is a Wigner distribution related by Fourier transform to a Generalized Transverse Momentum Dependent parton distribution (GTMD) (with zero skewness), 
i.e.\ a five-dimensional distribution, where even in the case of unpolarized scattering one has to take into account the directions of the transverse momentum and the impact parameter with respect to each other. This directional dependence has attracted much recent attention in large part due to the RHIC data for this process~\cite{STAR:2019wlg}, which shows a large $\cos 4 \phi$ angular modulation in the distribution of produced lepton pairs, where $\phi$ is the azimuthal angle between the transverse momentum of the lepton pair and of one of the leptons. This angular modulation is governed by the linear polarization of the photons, which arises analogously to the case of an ultrarelativistic electron that has an (equivalent) photon distribution that is highly, even maximally, linearly polarized, but now for charged ions (in case of the RHIC data Au ions). 
While the RHIC study focuses on one type of modulation, there will also be dependences on the azimuthal angles with respect to the impact parameter direction~\cite{Li:2019sin,Xiao:2020ddm,Shi:2024gex} or the event plane orientation, which are often taken to be equivalent. The most complete study to date is presented in Ref.~\cite{Shi:2024gex}, which considers the cross section differential in all three transverse vectors: the sum $\bm q_\sT$ and difference $\bm K_\sT$ of the transverse momenta of the pair of particles in the final state and the transverse impact parameter $\bm b_\sT$. This we will also consider here, finding agreement with previous studies, but we will present various new results as well. What we show in this paper is a GTMD description of the process including all unsuppressed angular modulations that appear in lepton and heavy-quark pair production, taking into account relevant mass terms, to obtain predictions that may help to further test the underlying formalism, e.g., in $\tau$ production~\cite{Shao:2023bga}.

The paper is organized as follows: in Section~\ref{sec correlator} we discuss the photon-photon GTMD correlator and its parameterization. In Section~\ref{sec process} the process of dilepton production in UPCs is examined, with a discussion of why this process can be expressed in terms of just two photon-photon GTMD correlators and why the skewness is (approximately) zero. In subsection \ref{subsec Delta space} we present the expressions for the differential cross section in $\bm\Delta_\sT$ space, which is most directly expressed in terms of GTMDs. 
However, as UPC experiments are unable to provide $\bm \Delta_\sT$ distributions, in contrast to its Fourier conjugate variable $\bm b_\sT$, in subsection \ref{subsec b space} we elaborate on the intricacies of Fourier transforming to $b\sT$ space with emphasis on how to project out the various azimuthal asymmetries that are most relevant for experimental studies of UPCs. In Section~\ref{sec: model and numerical result} we consider several models for the photon GTMDs in terms of form factors and compare them to RHIC data. We also present a host of predictions for the azimuthal modulations that are expected to arise and study their dependence on the mass of the produced lepton or heavy quark pair. We point out the most promising observables for experimental studies of this mass dependence. Section~\ref{sec conclusions} contains our conclusions. 
Finally, in Appendix \ref{sec: cross check} we compare some of our results to those already presented in the literature.

\section{The photon-photon GTMD correlator}\label{sec correlator}

\begin{figure}[t]
    \centering
    \includegraphics[width=0.7\linewidth, keepaspectratio, trim={5cm 21cm 3cm 5cm},clip]{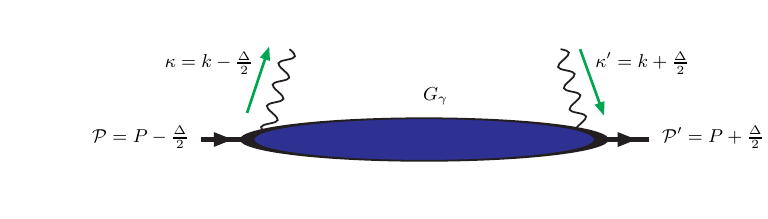}
\caption{\it The off-forward photon-photon correlator $G_\gamma^{\mu\nu}(x, \bm k_\sT,\xi, \bm \Delta_\sT)$. The momentum through the middle is $P - k$.} 
 \label{fig:corr}
\end{figure}

The photon-photon GTMD correlator describes the nucleus-to-photon transition and is defined, in analogy to the gluon-gluon correlator~\cite{Lorce:2013pza,More:2017zqp}, as the Fourier transform of a nonlocal, off-forward nuclear matrix element of two electromagnetic field strength tensors between two nuclear states with mass $M_A$, momenta ${\cal P}$ and ${\cal P}^\prime$ (${{\cal P}^2 = {{\cal P}^\prime}^2 = M_A^2}$), and average momentum $P \equiv ({\cal P} + {\cal P}^\prime)/2$. Between the two states a momentum $\Delta = {\cal P}^\prime - {\cal P}$ is transferred, exchanged by the emission and absorption of a photon. 
We define the momenta of these two photons as $\kappa$ and $\kappa^\prime$, with 
\begin{equation}
    k = (\kappa + \kappa^\prime)/2
\label{eq: GTMD photon mom sum}
\end{equation}
being their average and
\begin{equation}
    \kappa^\prime - \kappa = \Delta\, 
\label{eq: GTMD Delta constraint}
\end{equation}
their difference. Among the momentum assignments that satisfy Eqs.~\eqref{eq: GTMD photon mom sum} and~\eqref{eq: GTMD Delta constraint}, we choose the symmetric one, for which
\begin{equation}
    \kappa = k \, - \, \frac{\Delta}{2}\, , \quad \kappa^\prime = k \, + \, \frac{\Delta}{2}\, .
\label{eq: kappa definition}
\end{equation}
The corresponding pictorial representation of the correlator according to this choice is given in Fig.~\ref{fig:corr}.

A Sudakov decomposition of the momenta of the photons and nucleus can be performed by means of two light-like vectors $n$ and $\overline n$, satisfying $n^2 = \overline n^2 =0$ and $n\cdot \overline n =1$. The light-cone components of every vector $v$ are defined as $v^+ \equiv v \cdot \overline n$ and $v^-\equiv v\cdot n$, while perpendicular vectors $v_\sT$ always refer to the components of $v$ orthogonal to both $n$ and $\overline n$, with $v_\sT^2 = - \bm v_\sT^2$. 
Therefore, we have that
\begin{equation}
\begin{aligned}
P & = (P^+,P^-,\bm 0_\sT)\,,  \\
k & = (xP_N^+, k^-,\bm k_\sT)\,, \\
\Delta & = (-2 \xi P^+, 2\xi P^-, \bm \Delta_\sT)\,,
\label{eq:GTMDvectors}
\end{aligned}
\end{equation}
where $P_N$ is the nucleon momentum and the skewness parameter $\xi$ is defined as $2\xi = - \Delta^+/P^+ =- \Delta^+/(AP_N^+)$.
Note that although we consider the momentum fraction $x$ w.r.t.\ a nucleon in the nucleus, the off-forward momentum “kick” $\Delta$, and hence the skewness, applies to the nucleus as a whole. We do not consider the kick per nucleon, as it is not necessarily evenly distributed (also not to good approximation, since for instance $\Delta_\sT$ may be of the order of the average nucleon transverse momentum and may be taken by a single nucleon or by several or all of them, in contrast to $\Delta^+$ which is generally negligible w.r.t.\ $P_N^+$). Furthermore, from the relations $\Delta^2 = -4 \xi^2 P^2 - \bm \Delta_\sT^2$ and $\Delta^2 = 4 (M_{A}^2 - P^2)$, one obtains
\begin{align}
P^2 = 2 P^+ P^- = \frac{M_{A}^2 + \frac{1}{4}\,\bm \Delta_\sT^2}{1-\xi^2}\,.
\end{align}

Hence, for a photon emitted with average momentum $k$, we utilize the following definition of the correlator
\begin{equation}
G_\gamma^{\mu\nu}(x,\bm k_\sT, \xi, \bm \Delta_\sT) =
\frac{1}{x^2} \frac{1}{{\cal P}^+}
{\int}\frac{\d \zeta^-\,\d^2 \zeta_\sT}{(2\pi)^3}\
e^{i k\cdot \zeta}\,
\bigg\langle {\cal P}^\prime \bigg|\,\tr\left[\,F^{+\mu} \left (-\frac{\zeta}{2} \right )\, {\cal U}_{\left [-\frac{\zeta}{2}, \frac{\zeta}{2} \right]}
F^{+\nu}\left (\frac{\zeta}{2} \right)\,{\cal U}^\prime_{\left [\frac{\zeta}{2}, -\frac{\zeta}{2} \right]}
\right ] \bigg|{\cal P} \bigg\rangle\,\Bigg\rfloor_{\text{LF}}\,,
\label{eq:corr}
\end{equation}
where the nonlocality is restricted to the light-front LF: $\zeta \cdot \overline n \equiv \zeta^+ =0$. Note that we adopt a symmetric interval along the plus direction $[-\frac{\zeta}{2},\frac{\zeta}{2}]$, which for the general off-forward case is not the same as $[0,\zeta]$, because of the absence of translation invariance, but in the zero skewness case these are, in fact, the same. 
In contrast to the gluon case, the photon correlator does not need any gauge link to render it gauge invariant, which however does not mean 
they are not present or relevant. 
The gauge links sum up multiple photon exchange contributions that are in principle present. Although they are expected to play only a minor role in QED, because their effects are suppressed by powers of the small electromagnetic coupling $\alpha$, this suppression can be overcome in large field situations, like in heavy ion collisions where the coupling is enhanced by a high charge $Z$ (see also the discussion in Ref.~\cite{Mazurek:2021ahz}). The photon rescattering contributions can lead to e.g.\ the Landau-Pomeranchuk-Migdal effect \cite{Landau:1953um,Landau:1953gr,Migdal:1956tc,Baier:1996vi,Wiedemann:1999fq}, $p_T$-broadening, and process dependence (see e.g.\ Ref.~\cite{Boer:2015kxa} and references therein), but in the remainder of this paper we will simply leave them implicit, like in 
Eq.~(\ref{eq: photon distribution definition}), as they will play no further role here.

For a (heavy) nucleus that will generally not be polarized in a collider, the correlator can be parametrized in terms of the complex-valued GTMDs ${\cal F}_i^\gamma$, with $i=1,\dots,4$,\footnote{We note that ${\cal F}_4^\gamma$ is the analogue of the quark GTMD distribution $G_{1,1}$ of Refs.~\cite{Meissner:2009ww,Lorce:2011kd}, rather than $F_{1,4}$. The latter function is related to orbital angular momentum, whereas the former corresponds to a spin-orbit correlation. ${\cal F}_4^\gamma$ corresponds to circular photon polarization inside an unpolarized hadron, while ${\cal F}_{2}^\gamma$ and ${\cal F}_{3}^\gamma$ correspond to linear photon polarization.}
\begin{align}
G_\gamma^{\mu\nu}(x,\bm k_\sT ,\xi, \bm \Delta_\sT) & = \frac{1}{2x}\,\bigg \{-g_\sT^{\mu\nu}\,{\cal F}_1^\gamma (x,\bm k_\sT^2, \xi, \bm \Delta_\sT^2, \bm k_\sT \cdot \bm \Delta_\sT) \,+\, \frac{k_\sT^{\mu\nu}}{M_N^2}\, {\cal F}_2^\gamma (x,\bm k_\sT^2, \xi, \bm \Delta_\sT^2, \bm k_\sT \cdot \bm \Delta_\sT)\,  \nonumber \\  
    & \qquad  + \frac{\Delta_\sT^{\mu\nu}}{M_N^2}\, 
    {\cal F}_3^\gamma (x,\bm k_\sT^2, \xi, \bm \Delta_\sT^2, \bm k_\sT \cdot \bm \Delta_\sT) \, + \, \frac{k_\sT^{[\mu} \Delta_\sT^{\nu ]}}{M_N^2} \,  {\cal F}_4^\gamma (x,\bm k_\sT^2, \xi, \bm \Delta_\sT^2, \bm k_\sT \cdot \bm \Delta_\sT)  \bigg \}\,,
\label{eq: GTMDs}
\end{align}
where the symmetric transverse projector is given by
\begin{align}
g^{\mu\nu}_{\sT} = g^{\mu\nu} - n^{\mu} \overline n^{\nu} - n^{\nu}\overline n^{\mu}\, .
\end{align}
Moreover, the symmetric traceless tensors $\Delta_\sT^{\mu\nu}$ and $k_\sT^{\mu\nu}$ are defined as $a_\sT^{\mu\nu} = a_\sT^{\mu} a_\sT^{\nu} + \frac{1}{2}\, \bm a_\sT^2 \, g_\sT^{\mu\nu}$, and the square brackets denote antisymmetrization of the indices. On dimensional grounds, the tensor structures are divided by a mass squared. 
Since they indicate the photon polarization state that arises from the charged nucleons in the nucleus, we have chosen $M_N$ instead of $M_A$, as the latter might be interpreted as giving a suppression for large $A$. The adopted mass is just a matter of definition though, since the GTMDs in Eq.~\eqref{eq: GTMDs} are expected to scale as $M_N$ too.

The GTMDs ${\cal F}_i^\gamma$ are related to the leading-twist GPDs and TMDs for unpolarized hadrons by integrating over  $\bm k_\sT$ and by setting $\Delta=0$, respectively. Upon neglecting gauge links the GTMDs will be strictly real and for zero skewness can only depend on even powers of $\bm k_\sT \cdot \bm \Delta_\sT$ (see e.g.\ \cite{Meissner:2009ww,Boer:2018vdi}), as will be important later on.

\section{Dilepton production in nucleus-nucleus collisions}
\label{sec process}

\subsection{Theoretical framework}

We consider the reaction
\begin{align}
{\cal N}({\cal P}_1) + {\cal N}({\cal P}_2) \to \ell^+(K_1) + \ell^-(K_2) + X  \,,
\end{align}
where the four momenta of the particles are given within brackets, and the lepton pair in the final state is produced in an ultraperipheral collision of two highly charged nuclei. We focus on the specific kinematic configuration where the leptons are almost back to back in the plane perpendicular to the direction of the initial nuclei. 
The final state $X$ consists mostly of the same two nuclei, as the energy carried away by the photons is small. It is however possible for the nuclei to mutually excite each other through photon exchange, leading to subsequent emission of one or more neutrons along both beam directions. The cross sections for emission of one or more neutrons (labeled $XnXn$)\footnote{In the literature, $Xn$ usually indicates all the other neutron emission channels. Hence, if an experiment does not measure specific neutron channels, $Xn$ refers to one or more emissions, if $1n$ is measured $Xn$ refers to two or more, etc. However, in this paper we only compare to STAR data~\cite{STAR:2019wlg}, where $Xn$ corresponds to the “one or more neutrons” case.}
and exactly one neutron on each side (labeled $1n1n$) are respectively about $10$ and $100$ times smaller than the untagged cross section which is thus mostly $0n0n$ \cite{Baltz:2002pp}.

\begin{figure}[t]
    \centering
    \hspace*{-0.7cm}
    \includegraphics[width=0.49\linewidth, keepaspectratio, trim={5cm 21cm 2.5cm 1cm},clip]{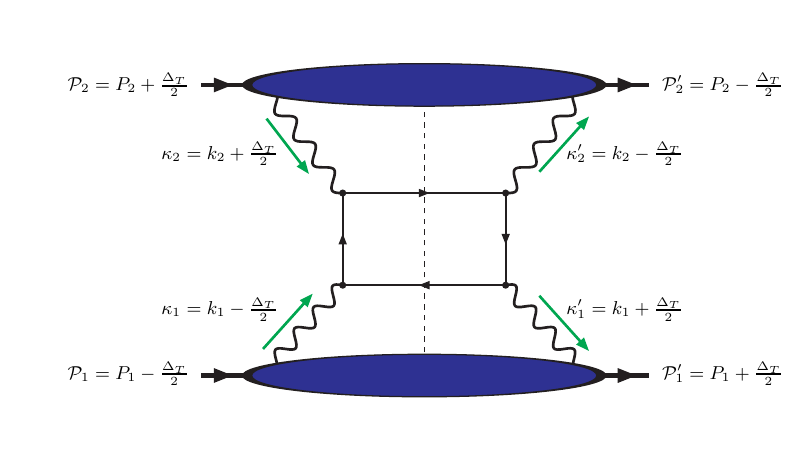}\hspace*{0.2cm}
    \includegraphics[width=0.49\linewidth, keepaspectratio, trim={5cm 21cm 2.5cm 1cm},clip]{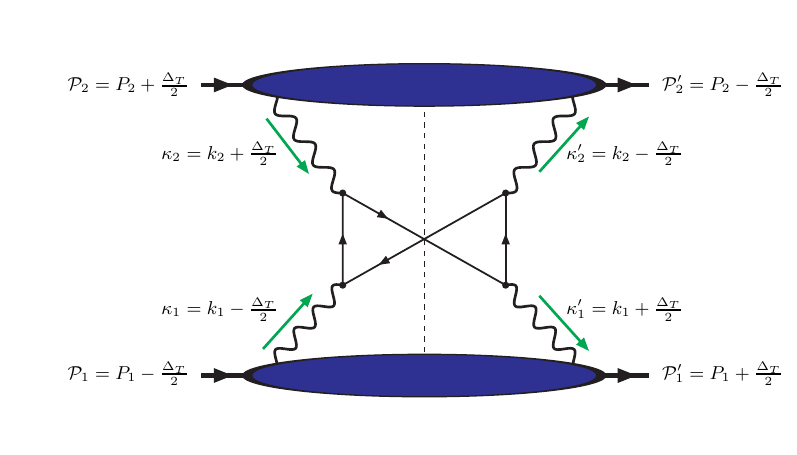}
    \caption{\it Representative cut diagrams for the process ${\cal N}{\cal N}\to \ell^+\ell^-X$. The other two diagrams, in which the directions of the arrows in the fermionic lines are reversed, are not shown. Momentum conservation implies $\Delta_\sT \equiv \Delta_{1\sT} = -\Delta_{2\sT}$.} 
    \label{fig:fd-cs}
\end{figure}

The dominant lepton-pair production channel is the photon-photon fusion subprocess
\begin{align}
\gamma({\kappa}_1) + \gamma(\kappa_2) \to \ell^+(K_1) + \ell^-(K_2)\,,
\end{align}
as described by the relevant Feynman diagrams in Fig.~\ref{fig:fd-cs}. The photons can have different momenta in the amplitude ($\kappa_{1},~\kappa_{2}$) compared to its complex conjugated ($\kappa_{1}^\prime,~\kappa_{2}^\prime$), with the differences determined by ($\Delta_1,~\Delta_2$), respectively.
However, the total momentum of the final state does not vary between the amplitude and its complex conjugate, which leads to the constraint
\begin{align}
\kappa_1^\prime + \kappa_2^\prime = \kappa_1 + \kappa_2 = K_1 + K_2\,, 
\label{eq:mom-cons}
\end{align}
and
\begin{align}
\Delta \equiv \Delta_1 = -\Delta_2\,. 
\label{eq: DeltaT constraint}
\end{align}

Moreover, due to momentum conservation, the momentum transfer is restricted to the transverse plane. To see this, we can perform the Sudakov decomposition of the different momenta using the same light-like vectors $n$ and $\overline n$ introduced in the previous section.
Hence, according to Eq.~\eqref{eq:GTMDvectors}, we have that the average nuclear momenta are
\begin{align}
{P}_1^\mu & = {P}_1^+n^\mu + \frac{P_1^2}{2{P}_1^+}\, \overline n^\mu\,, \qquad
{P}_2^\mu  = \frac{P_2^2}{2{P}_2^-}\, n^\mu  + {P}_2^- \overline n^\mu\,,
\end{align}
and the photon momenta can be expressed as
\begin{align}
\kappa_{1}^\mu & = (x_1 + A\xi) {P}_{1N}^+ n^\mu + \frac{\bm \kappa_{1\sT}^2}{2 (x_1 + A\xi) {P}_{1N}^+}\, \overline n^\mu +  \kappa_{1\sT}^\mu \approx (x_1 + A\xi) {P}_{1N}^+ n^\mu +   \kappa_{1\sT}^\mu\, , \\
\kappa_{2}^\mu & = \frac{\bm \kappa_{2\sT}^2}{2 (x_2 - A\xi) {P}_{2N}^-}\, n^\mu +  (x_2 - A\xi) {P}_{2N}^- \overline n^\mu +\kappa_{2\sT}^\mu\approx \, (x_2 - A\xi) {P}_{2N}^- \overline n^\mu +\kappa_{2\sT}^\mu,
\end{align}
where $P_{iN}$, with $i=1,2$, is the momentum of a nucleon $N$ inside the nucleus ${\cal N}(P_i)$. Similarly, one can decompose $\kappa_{1}^{\prime}$ and $\kappa_{2}^{\prime}$ to obtain  that 
\begin{align}
\Delta_1^\mu \approx -2 \xi P_1^+ n^\mu + \Delta_\sT^\mu\, , \quad \Delta_2^\mu\, \approx \,2\xi P_1^- \overline n^\mu - \Delta_\sT^\mu\,, 
\end{align}
which can only satisfy Eq.~\eqref{eq: DeltaT constraint} for $\xi=0$. In other words, $\xi$ is as suppressed as the “$-$” components of $\kappa_1$ and $\kappa_1^\prime$ and the “$+$” components of $\kappa_2$ and $\kappa_2^\prime$.
Therefore, as already anticipated, in this approximation the momentum is transferred solely along the transverse direction with $\Delta = \Delta_\sT$, justifying the momentum assignments in Fig.~\ref{fig:fd-cs}. 

Next we consider the momenta of the leptons in the final state for which
\begin{equation}
    K_1^\mu = \frac{M_{1\sT}}{\sqrt{2}}\left( \e^{y_1}\, n^\mu + \e^{-y_1}\, \overline n^\mu \right) + K_{1\sT}^\mu\,, \qquad K_2^\mu = \frac{M_{2\sT}}{\sqrt{2}}\left( \e^{y_2}\, n^\mu + \e^{-y_2}\, \overline n^\mu \right) + K_{2\sT}^\mu\, ,
\end{equation}
with $y_1$ and $y_2$ being their rapidities. Here we introduced the transverse masses ${M_{i\sT} = \sqrt{M_\ell^2 + \bm K_{i\sT}^2}}$, with $i = 1,2$ and where $M_\ell$ is the lepton mass. Besides, we define the sum and difference of the outgoing particle transverse momenta as $q_\sT = K_{1\sT} + K_{2\sT}$ and $K_\sT = (K_{1\sT} - K_{2\sT})/2$, respectively, with $\vert q_\sT\vert \ll \vert K_\sT\vert$. In this specific configuration, called the correlation limit, the approximations
$K_{1\sT} \approx K_{\sT}$ and $K_{2\sT} \approx -K_{\sT}$ hold, and consequently
$M_\sT\approx M_{1\sT}\approx M_{2\sT}$.
Specifying the different light-cone components of Eq.~(\ref{eq:mom-cons}),  and using that $\xi \approx 0$, we find
\begin{align}
x_1\, P_{1N}^+ & \approx K_1^+ + K_2^+\,, \nonumber \\
x_2\, P_{2N}^- & \approx K_1^- + K_2^-\,, \nonumber \\
\kappa_{1\sT}^\prime + \kappa_{2\sT}^\prime & = \kappa_{1\sT} + \kappa_{2\sT}  = k_{1\sT} + k_{2\sT} = q_\sT\, .\nonumber 
\end{align}
As a result, the photon momentum fractions in the nucleon-nucleon  
c.m.~frame are simply given by
\begin{equation}
    x_1 = \frac{M_{1\sT}\, \e^{y_1} + M_{2\sT}\, \e^{y_2}}{\sqrt{s_{NN}}} \approx \frac{M_{\sT}}{\sqrt{s_{NN}}} \left( \e^{y_1} + \e^{y_2} \right)\, , \qquad 
    x_2 = \frac{M_{1\sT}\, \e^{-y_1} + M_{2\sT}\, \e^{-y_2}}{\sqrt{s_{NN}}} \approx \frac{M_{\sT}}{\sqrt{s_{NN}}} \left( \e^{-y_1} + \e^{-y_2} \right)\, .
\label{eq:x1-x2}
\end{equation}
Furthermore, the final results can be expressed in terms of the following energy fractions of the final leptons
\begin{equation}
z_1 \equiv \frac{K_1 \cdot \kappa_1}{\kappa_1 \cdot \kappa_2} = \frac{1}{1 + e^{y_1 - y_2}}\, ,\qquad
 z_2 \equiv \frac{K_2 \cdot \kappa_1}{\kappa_1 \cdot \kappa_2} = \frac{1}{1 + e^{y_2 - y_1}}\,.
\end{equation}
Due to the momentum conservation in Eq.~\eqref{eq:mom-cons},
one has $z_1 + z_2 = 1$, and we therefore define $z \equiv z_1$, such that $z_2 = 1 - z$. 
Thus, the partonic Mandelstam variables are written in terms of $z$, $M_\ell$ and $K_{\sT}$ according to
\begin{equation}
\begin{aligned}
    \hat s & = (\kappa_1 + \kappa_2)^2 \approx \frac{M_{\sT}^2}{z\, (1 - z)}\, , \\
    \hat t & = (\kappa_1 - K_1)^2 \approx M_{\ell}^2 - \frac{M_{\sT}^2}{1 - z}\, , \\
    \hat u & = (\kappa_1 - K_2)^2 \approx M_{\ell}^2 - \frac{M_{\sT}^2}{z}\, .
\end{aligned}
\end{equation} 

The cross section in $\Delta_\sT$-space is expressed in terms of GTMDs at zero skewness according to
\begin{equation}
    \frac{\d\sigma}{\d{\rm PS}\, \d^2 \Delta_\sT} = \frac{1}{16\pi^2 s_{NN}^2} \int \d^2 \bm k_{1\sT}\, \d^2 \bm k_{2\sT}\, G_\gamma^{\mu\nu}(x_1 {,}\bm k_{1 \sT}, \bm \Delta_\sT)\, G_\gamma^{\rho\sigma}(x_2 {,}\bm k_{2 \sT}, -\bm \Delta_\sT)\, {\cal M}_{\mu\rho} \,{\cal M}^{*}_{\nu\sigma}\, \delta^2(\bm k_{1 \sT} + \bm k_{2 \sT} -\bm q_\sT )\, , 
 \label{eq:cs}
\end{equation}
where $\d{\rm PS} = \d y_1 \, \d y_2\, \d^2 \bm K_\sT\, \d^2 \bm q_\sT$ and ${\cal{M}}$ is the hard scattering amplitude for the process $\gamma \gamma\to \ell^+\ell^-$, which is derived from the cut diagrams given in Fig.~\ref{fig:fd-cs}.
One may wonder whether these are actual cut diagrams, since the amplitudes on the right of the cut are not simple conjugates of the ones on the left. However, the photons in the nucleus (in the equivalent photon approximation as in Ref.~\cite{Vidovic:1992ik}) are considered not to be in a momentum eigenstate. In this way there can be incoming photons with different momenta in the amplitude compared to the complex conjugated one, and, actually, the nonzero contributions to the cross section are only given by such interference terms. By not considering momentum eigenstates, the cross section is already nonzero at the level of two GTMDs, rather than four (i.e.\ one for each nucleus in both amplitude and conjugate amplitude), and Fig.~\ref{fig:fd-cs} refers indeed to cut diagrams.

Before elaborating more on this important point, we would like to emphasize another, equally important, observation:
the variable $\bm \Delta_\sT$ is not directly observable in UPC experiments (unless one would measure the momenta of the recoiling nuclei precisely), whereas its Fourier conjugate variable, the impact parameter $\bm b_\sT$, can be determined, although not event by event.
In particular, the angle of $\bm b_\sT$ can be approximated with the event plane orientation, whilst its length is estimated by means of Monte Carlo or Machine Learning simulations~\cite{Roland:2014jsa, Mallick:2021wop} which evaluate the correlations between $b_\sT$ and other observables, like the multiplicity of charged particles. The events with different impact parameters are usually binned in ranges of centrality, necessitating the integration over the corresponding $b_\sT$ range.
The cross section in $b_\sT$-space can formally be obtained from Eq.~\eqref{eq:cs} via the Fourier transform
\begin{equation}
    \frac{\d \sigma}{\d {\rm PS}\,  \d^2 \bm b_\sT} = \int \frac{\d^2\bm \Delta_\sT}{(2\pi)^2} \, \e^{i \bm b_\sT \cdot \bm \Delta_\sT}\, \frac{\d\sigma}{\d{\rm PS}\, \d^2\bm \Delta_\sT}\, .
\end{equation}
The Fourier transform turns a product of $\Delta_\sT$-dependent expressions into a convolution of $b_\sT$-dependent terms:  
\begin{multline}
    \int \frac{\d^2\bm{\Delta}_\sT}{(2\pi)^2}\, \e^{i \bm{b}_\sT \cdot \bm{\Delta}_\sT} \; {\cal F}_i^\gamma(x_1, \bm{k}_{1\sT}, \bm{\Delta}_\sT)\, {\cal F}_j^\gamma(x_2, \bm{k}_{2\sT}, -\bm{\Delta}_\sT) \cdots \\
    = \int \d^2 \bm b_{1\sT} \d^2\bm  b_{2\sT}\, \delta^2(\bm{b}_\sT -\bm{b}_{1\sT} + \bm{b}_{2\sT}) \, f^i_{\gamma/A}(x_1, \bm{k}_{1\sT}, \bm{b}_{1\sT}) f^j_{\gamma/A}(x_2, \bm{k}_{2\sT}, \bm{b}_{2\sT}) \cdots \, ,
\label{eq:csFT}
\end{multline}
where the dots indicate all the terms that coincide on the left- and right-hand sides of Eq.~\eqref{eq:csFT}. Although these expressions apply formally, in practice one has to deal with some subtleties. One of them concerns the required lower limits on the $b_{i\sT}$ integration regions \cite{Mazurek:2021ahz} set by the radii of the nuclei, which excludes contributions of photons inside the nucleus. In Sec.~\ref{sec: model and numerical result} we will explicitly deal with $b_{\min}$.
Another subtlety with Eq.~\eqref{eq:csFT} has to do with the center of the nuclei with respect to which one defines the impact parameter. For this, one has to carefully consider wave packets in the same way as done for impact parameter dependent GPDs $q(x,\bm{b}_\sT)$ \cite{Burkardt:2000za,Burkardt:2002hr,Burkardt:2002ks,Diehl:2002he}, which are diagonal matrix elements of the form $\langle P^+, \bm{R}_\sT=0| O(\bm{b}_\sT)| P^+, \bm{R}_\sT=0 \rangle$, with $\bm b_\sT$ being the impact parameter measured with respect to the transverse center of longitudinal momentum $\bm{R}_\sT\equiv \sum_i x_i \bm{r}_{\sT i}$~\cite{Soper:1976jc,Burkardt:2000za}. Since the states $| P^+, \bm{R}_\sT=0 \rangle$ are not momentum eigenstates, $q(x,\bm{b}_\sT)$ and consequently its Fourier conjugate which is the standard GPD \cite{Burkardt:2000za,Burkardt:2002hr,Burkardt:2002ks,Diehl:2002he} correspond to off-forward matrix elements of the form $\langle P'| O(\bm{b}_\sT) | P\rangle$ (for more details see e.g.\ Ref.~\cite{Boer:2021upt}). This applies to the GTMDs in precisely the same way, meaning that the off-forwardness implies that one is not dealing with momentum eigenstates. 

More explicitly, Eq.~\eqref{eq:csFT} can be seen as connecting the photon distributions in $\bm b_\sT$-space to the GTMDs introduced in Eq.~\eqref{eq:cs}.
If we (temporarily) ignore the $k_\sT$ dependence, the photon distribution as a function of the impact parameter $\bm{b}_\sT$ in a nucleus localized around its transverse center of longitudinal momentum $\bm{R}_\sT$ is given by  
\begin{align}
f_{\gamma/A}(x,\bm{b}_\sT) = \frac{1}{x (P^+)^2} \int \frac{\d \lambda}{2\pi} \, \e^{i x \lambda}\, 
\langle {P^+, \bm{R}_\sT = 0} \vert\;  
F^{+\mu}(-\frac{\lambda}{2}n+{\bm{b}_\sT}) F^{+}{}_{\mu}(\frac{\lambda}{2}n+{\bm{b}_\sT}) \; 
\vert {P^+, \bm{R}_\sT = 0} \rangle. 
\label{nomegab}
\end{align}
Here $\vert {P^+, \bm{R}_\sT = 0} \rangle$ denotes the normalized nuclear state localized in the transverse spatial direction \cite{Diehl:2002he,Burkardt:2002ks}:
\begin{align}
\vert {P^+, \bm{R}_\sT = 0} \rangle = {\cal N} \int \frac{\d^2 \bm{P}_\sT}{(2\pi)^2} \Phi(\bm{P}_\sT) |P^+,\bm{P}_\sT \rangle,
\end{align}
for some wave packet $\Phi(\bm{P}_\sT)$. If this wave packet is sufficiently localized in the transverse spatial direction, i.e.\ $\Phi(\bm{P}_\sT+\bm{\Delta}_\sT) \approx \Phi(\bm{P}_\sT)$, one can relate the number of photons to an off-forward distribution ${\cal F}_1^\gamma$:
\begin{align}
f_{\gamma/A}(x,\bm{b}_\sT) & = |{\cal N}|^2 \int \frac{\d \lambda \, \e^{i x \lambda}}{2\pi x(P^+)^2}\int \frac{\d^2\bm{P}_\sT}{(2\pi)^2} \frac{\d^2\bm{P}^\prime_\sT}{(2\pi)^2} \, \Phi^*(\bm{P}^\prime_\sT) \Phi(\bm{P}_\sT)   
\langle {P^+, \bm{P}^\prime_\sT} \vert\; F^{+\mu}(-\frac{\lambda}{2}n+{\bm{b}_\sT}) F^{+}{}_{\mu}(\frac{\lambda}{2}n+{\bm{b}_\sT}) \; \vert {P^+, \bm{P}_\sT} \rangle\nonumber\\
 & = |{\cal N}|^2 \int \frac{\d \lambda\, \e^{i x \lambda}}{2\pi x (P^+)^2}\int \frac{\d^2\bm{P}_\sT}{(2\pi)^2} \frac{\d^2\bm{P}^\prime_\sT}{(2\pi)^2} \, \Phi^*(\bm{P}^\prime_\sT) \Phi(\bm{P}_\sT)  \e^{-i\bm{b}_\sT \cdot \bm{\Delta}_\sT} \langle {P^+, \bm{P}^\prime_\sT} \vert\;  
F^{+\mu}(-\frac{\lambda}{2}n) F^{+}{}_{\mu}(\frac{\lambda}{2}n) \; \vert {P^+, \bm{P}_\sT} \rangle\nonumber\\
& \approx |{\cal N}|^2 \int \frac{\d^2\bm{P}_\sT}{(2\pi)^2} |\Phi(\bm{P}_\sT)|^2 \frac{\d^2\bm{\Delta}_\sT}{(2\pi)^2} \, \e^{-i\bm{b}_\sT \cdot \bm{\Delta}_\sT} {\cal F}_1^\gamma(x,\bm{\Delta}^2_\sT) \nonumber\\
& = \int \frac{\d^2\bm{\Delta}_\sT}{(2\pi)^2} \, \e^{-i\bm{b}_\sT \cdot \bm{\Delta}_\sT} {\cal F}_1^\gamma(x,\bm{\Delta}^2_\sT).
\end{align}
This can be generalized to the transverse momentum dependent distribution ${\cal F}_1^\gamma(x,\bm{k}_\sT,\bm{\Delta})$, which is the GTMD of Eq.~\eqref{eq: GTMDs} for zero skewness.
In this way one generalizes the cross section expression $\sigma_{AA}$ in Eq.~\eqref{eq: photon-photon cross section} 
by going from integrals over the one-dimensional PDFs $f_{\gamma/A}(x)$ to GPDs and GTMDs. A full derivation of the impact parameter dependent cross section in terms of wave packets and off-forward matrix elements, as well as the connection to the TMD description of the $b_\sT$-integrated cross section was presented in Ref.~\cite{Wang:2021kxm}.

\subsection{Expressions in \texorpdfstring{$\Delta_\sT$}{Δ{\textunderscore}T}-space}
\label{subsec Delta space}

By employing the correlator in Eq.~\eqref{eq:corr}, we can express Eq.~\eqref{eq:cs} in the following form 
\begin{align}
    \frac{\d\sigma}{\d {\rm PS}\,\d^2 \bm \Delta_\sT} = \frac{\alpha^2}{s_{NN} M_\sT^2}\,& \bigg[ 
    F^0 + F^{\cos 2\phi_{q\Delta}} \cos 2\phi_{q\Delta} + F^{\sin 2\phi_{q\Delta}} \sin 2\phi_{q\Delta} \nonumber \\
    & \phantom{=} + F^{\cos 2 \phi_{qK} } \cos 2\phi_{qK} + F^{\sin 2\phi_{qK} } \sin 2\phi_{qK} + F^{\cos 2\phi_{\Delta K}} \cos 2\phi_{\Delta K}  \nonumber \\
    & \phantom{=}+ F^{\cos 4\phi_{qK}} \cos 4\phi_{qK} + F^{\sin 4\phi_{qK}} \sin 4\phi_{qK} + F^{\cos 4\phi_{\Delta K}} \, \cos 4\phi_{\Delta K} \nonumber \\
    & \phantom{=} + F^{\cos 2(\phi_{qK} + \phi_{\Delta K})} \cos 2(\phi_{qK} + \phi_{\Delta K}) + F^{\sin 2(\phi_{qK} + \phi_{\Delta K})} \, \sin 2(\phi_{qK} + \phi_{\Delta K})  \bigg]\,,
\label{eq: cs dileqepton DeltaT-space}
\end{align}  
where we have introduced the notation $\phi_{ab} = \phi_a - \phi_b$, and where
\begin{subequations}
\begin{equation}
\begin{aligned}
F^0 & = 
    2 \left[ z^2 + (1 - z)^2 + 4 z\, (1 - z)\, \frac{M_\ell^2}{M_\sT^2} \left(1 - \frac{M_\ell^2}{M_\sT^2} \right )\right]{\cal C}[{\cal F}_1^\gamma\, {\cal F}_1^\gamma] \\ 
    & \phantom{=} 
    - z\, (1 - z) \frac{M_\ell^4}{M_\sT^4} \left({\cal C}[w_0^{22}\, {\cal F}_2^\gamma\, {\cal F}_2^\gamma] + \frac{\bm \Delta_\sT^4}{M_N^4} {\cal C}[{\cal F}_3^\gamma\, {\cal F}_3^\gamma] \right)
    + \left[ z^2 + (1 - z)^2\right] \left( 1 - 2\frac{M_\ell^2}{M_\sT^2} \right) \frac{\bm \Delta_\sT^2}{M_N^2}\, {\cal C}[w_0^{44}\, {\cal F}_4^\gamma\, {\cal F}_4^\gamma]\,, \\
 F^{\cos 2\phi_{q \Delta}} & = 
    - \frac{\bm \Delta_\sT^2}{M_N^2} \Bigg\{ \left[ z^2 + (1 - z)^2 \right] \left( 1 - 2 \frac{M_\ell^2}{M_\sT^2} \right) {\cal C} [w_{c2}^{44} {\cal F}_4^\gamma {\cal F}_4^\gamma] + z\, (1 - z)\, \frac{M_\ell^4}{M_\sT^4}\, {\cal C}[w_{c2}^{23}\, {\cal F}_2^\gamma \, {\cal F}_3^\gamma + w^{32}_{c2}\, {\cal F}_3^\gamma \, {\cal F}_2^\gamma]\Bigg\}\,, \\
 F^{\sin 2\phi_{q \Delta}} & = 
    - \frac{\bm \Delta_\sT^2}{M_N^2} \Bigg\{ \left[ z^2 + (1 - z)^2 \right] \left( 1 - 2 \frac{M_\ell^2}{M_\sT^2} \right) {\cal C} [w_{s2}^{44} {\cal F}_4^\gamma {\cal F}_4^\gamma] + z\, (1 - z)\, \frac{M_\ell^4}{M_\sT^4}\, {\cal C}[w_{s2}^{23}\, {\cal F}_2^\gamma \, {\cal F}_3^\gamma + w^{32}_{s2}\, {\cal F}_3^\gamma \, {\cal F}_2^\gamma] \Bigg\}\, ,
\label{eq: structure functions explicit - 0phiK}
\end{aligned}
\end{equation}
\begin{equation}
\begin{aligned}
F^{\cos 2\phi_{qK}} & = 
    4\, z\, (1 - z) \,\frac{M_\ell^2}{M_\sT^2} \left (1 -\frac{M_\ell^2}{M_\sT^2} \right ) {\cal C}[w_{c2}^{12}\, {\cal F}_1^\gamma \, {\cal F}_2^\gamma + w_{c2}^{21}\, {\cal F}_2^\gamma\, {\cal F}_1^\gamma]\,, \\
F^{\sin 2\phi_{qK}} & = 
    4\, z\, (1 - z) \,\frac{M_\ell^2}{M_\sT^2} \left( 1 - \frac{M_\ell^2}{M_\sT^2} \right) {\cal C}[w_{s2}^{12}\, {\cal F}_1^\gamma\, {\cal F}_2^\gamma + w_{s2}^{21}\, {\cal F}_2^\gamma\, {\cal F}_1^\gamma] \,, \\
F^{\cos 2\phi_{\Delta K}} & = 
    4\, z\,(1 - z)\, \frac{M_\ell^2}{M_\sT^2}\, \left( 1 -\frac{M_\ell^2}{M_\sT^2} \right) \frac{\bm \Delta_\sT^2}{M_N^2} \,{\cal C}[{\cal F}_1^\gamma \,{\cal F}_3^\gamma + {\cal F}_3^\gamma \, {\cal F}_1^\gamma]\,,
\label{eq: structure functions explicit - 2phiK}
\end{aligned}
\end{equation}
\begin{equation}
\begin{aligned}
F^{\cos 4\phi_{qK}} & = 
    - z\, (1 - z)\, \bigg(1 - \frac{M_\ell^2}{M_\sT^2} \bigg)^2 {\cal C}[w_{c4}^{22}\, {\cal F}_2^\gamma \, {\cal F}_2^\gamma]\,, \\
F^{\sin 4 \phi_{qK}} & = 
    - z\, (1 - z)\, \bigg(1 - \frac{M_\ell^2}{M_\sT^2} \bigg)^2 {\cal C}[w_{s4}^{22}\, {\cal F}_2^\gamma \, {\cal F}_2^\gamma]\,, \\
 F^{\cos 4\phi_{\Delta K}} & = 
    -z\, (1 - z)\, \left( 1 -\frac{M_\ell^2}{M_\sT^2} \right)^2 \frac{\bm \Delta_\sT^4}{M_N^4}\, {\cal C}[{\cal F}_3^\gamma \, {\cal F}_3^\gamma]\, , \\ 
F^{\cos 2(\phi_{q K} + \phi_{\Delta K})} & = 
    - z\, (1 - z)\, \bigg(1 - \frac{M_\ell^2}{M_\sT^2} \bigg)^2 \frac{\bm \Delta_\sT^2}{M_N^2}\, {\cal C} [w_{c2}^{23} \, {\cal F}_2^\gamma \, {\cal F}_3^\gamma + {w}^{32}_{c2}\, {\cal F}_3^\gamma \, {\cal F}_2^\gamma ]\, , \\
F^{\sin 2(\phi_{q K} + \phi_{\Delta K})} & = 
    - z\, (1 - z)\, \bigg(1 - \frac{M_\ell^2}{M_\sT^2} \bigg)^2\, \frac{\bm \Delta_\sT^2}{M_N^2}\, {\cal C}[ w_{s2}^{23} \, {\cal F}_2^\gamma \, {\cal F}_3^\gamma + {w}^{32}_{s2}\, {\cal F}_3^\gamma \, {\cal F}_2^\gamma ]\,.
\label{eq: structure functions explicit - 4phiK}
\end{aligned}
\end{equation}
\label{eq: structure functions explicit}
\end{subequations}
In the above equations we have introduced the convolutions of two GTMDs, 
\begin{align}
    & {\cal{C}}[w\, {\cal F}_i\, {\cal F}_j](\bm q_\sT^2, \bm \Delta_\sT^2, \bm \Delta_\sT \cdot \bm q_\sT) \nonumber \\ & \qquad \equiv \int \d^2  k_{1\sT}\,\d^{2}k_{2\sT}\,
    \delta^{2}(\bm k_{1\sT} + \bm k_{2\sT} - \bm q_{\sT})\, w(\bm k_{1\sT},\bm k_{2\sT}, \bm \Delta_\sT)\, {\cal F}_i (x_1, \bm k_{1\sT}^2, \bm \Delta_\sT^2, \bm k_{1\sT} \cdot \bm \Delta_\sT)\, {\cal F}_j (x_2, \bm k_{2\sT}^2, \bm \Delta_\sT^2, - \bm k_{2\sT} \cdot \bm \Delta_\sT)\, ,
\label{eq: GTMD convolution definition}
\end{align}
where $x_1$ and $x_2$ are given in Eq.~\eqref{eq:x1-x2}
and $w$ is a transverse momentum dependent weight function. In order to provide compact expressions for the weight functions, we introduce the following two orthogonal unit vectors spanning the transverse plane~\cite{Boer:2003ya}
\begin{align}
\hat{\bm h}  = \frac{\bm q_\sT}{\vert \bm q_\sT\vert} = (\cos\phi_q,\ \sin\phi_q) \,, \qquad 
\hat {\bm g} & = (\sin\phi_q,\ -\cos\phi_q) \,, 
\end{align}
with the latter defined as 
\begin{align}
\hat{g}^{i} & = \epsilon_\sT^{ij} \, h^j \equiv \epsilon^{-+ij}h^j = \epsilon^{0ij3}\, h^j\,, 
\end{align} 
such that for any three-vector $\bm a$, 
$\hat {\bm g}  \cdot \bm a = (\bm a\times \hat{\bm h})_z$. Hence, if $\hat {\bm a} = (\cos\phi_a,~\sin\phi_a)$ then $\hat {\bm h} \cdot \hat{\bm a} = \cos(\phi_q - \phi_a)$ and $\hat {\bm g} \cdot \hat{\bm a} = \sin(\phi_q - \phi_a)$. The weight functions in Eqs.~\eqref{eq: structure functions explicit - 0phiK}-\eqref{eq: structure functions explicit - 4phiK} can be therefore written as
\begin{subequations}
\begin{equation}
\begin{aligned}
w_0^{22} & = 
    \frac{1}{M_N^4}\, \left[ 2\, (\bm k_{1\sT} \cdot \bm k_{2T})^2 - \bm k_{1\sT}^2 \bm k_{2\sT}^2 \right] = \frac{\bm k^2_{1\sT} \bm k^2_{2\sT}}{M_N^4}\, \cos 2\phi_{12}\,, \\
w_0^{44}& 
    = \frac{\bm k_{1\sT} \cdot \bm k_{2\sT}}{M_N^2} = \frac{|\bm k_{1\sT}| |\bm k_{2\sT}|}{M_N^2}\, \cos\phi_{12} \,, 
\label{eq: weights - c0}
\end{aligned}
\end{equation}
\begin{equation}
\begin{aligned}
w_{c2}^{44} & = 
    \frac{1}{M_N^2}\, \left[ 2\,  ( \hat{\bm h} \cdot \bm k_{1\sT})( \hat{\bm h} \cdot \bm k_{2\sT}) - ( \bm k_{1\sT} \cdot \bm k_{2\sT}) \right] = \frac{\vert \bm k_{1\sT} \vert\vert \bm k_{2\sT} \vert }{M_N^2} \cos(\phi_{q1} + \phi_{q2}) \, , \\
w_{s2}^{44} & = 
   \frac{(\hat{\bm h} \cdot \bm k_{1\sT})\, (\hat{\bm g} \cdot \bm k_{2\sT}) + (\hat{\bm h} \cdot \bm k_{2\sT})\, (\hat{\bm g} \cdot \bm k_{1\sT})}{M_N^2} = \frac{|\bm k_{1\sT}| |\bm k_{2\sT}|}{M_N^2} \, \sin(\phi_{q1} + \phi_{q2})\, , \\
w_{c2}^{23} & = 
    \frac{1}{M_N^2}\, \left[ 2\, {(\hat{\bm h} \cdot \bm k_{1\sT})^2} - \bm k_{1\sT}^2 \right] = \frac{\bm k_{1\sT}^2}{M_N^2} \cos 2\phi_{q1}\, ,\\
w_{s2}^{23} & = 
    \frac{2}{M_N^2}\, (\hat{\bm h} \cdot {\bm k_{1\sT}})\, (\hat{\bm g}\cdot {\bm k_{1\sT}}) = \frac{\bm k_{1\sT}^2}{M_N^2} \sin 2\phi_{q1}\,, \\
w_{c2}^{32} & = 
    \frac{1}{M_N^2}\, \left[ 2\, ( \hat{\bm h} \cdot \bm k_{2\sT})^2 -\bm k_{2\sT}^2 \right] = \frac{\bm k_{2\sT}^2}{M_N^2} \cos 2 \phi_{q2}\, ,\\
w_{s2}^{32} & = 
    \frac{2}{M_N^2}\, (\hat{\bm h} \cdot {\bm k_{2\sT}})\, (\hat{\bm g} \cdot {\bm k_{2\sT}}) = \frac{\bm k_{2\sT}^2}{M_N^2} \sin 2\phi_{q2}\,,
\label{eq: weights - c2}
\end{aligned}
\end{equation}
\begin{equation}
\begin{aligned}
w_{c4}^{22} & = 
    \frac{1}{M_N^4}\, \left \{ 2 \left[ 2 \, (\hat {\bm h} \cdot \bm k_{1\sT})\, (\hat {\bm h} \cdot \bm k_{2T}) - (\bm k_{1\sT} \cdot \bm k_{2\sT}) \right]^2 - \bm k_{1\sT}^2  \bm k_{2\sT}^2 \right \} = \frac{\bm k_{1\sT}^2 \bm k_{2\sT}^2}{M_N^4} \, \cos2(\phi_{q1} + \phi_{q2}) \,, \\
w_{s4}^{22} & = 
   \frac{2}{M_N^4}\, \left[ 2\, (\hat{\bm h} \cdot \bm k_{1\sT})\, (\hat{\bm h} \cdot \bm k_{2\sT}) -  (\bm k_{1\sT} \cdot \bm k_{2\sT} )\right]\left[ (\hat{\bm h} \cdot \bm k_{1\sT})\, (\hat{\bm g} \cdot \bm k_{2\sT}) + (\hat{\bm h} \cdot \bm k_{2\sT})\, (\hat{\bm g} \cdot \bm k_{1\sT}) \right]\\
    & = \frac{\bm k_{1\sT}^2 \bm k_{2\sT}^2}{M_N^4} \, \sin2(\phi_{q1} + \phi_{q2}) \, ,
\label{eq: weights - c4}
\end{aligned}
\end{equation}
and with the following relations holding
\begin{equation}
w_{c2}^{12} = w_{c2}^{32} \, , \quad
w_{c2}^{21} = w_{c2}^{23} \, , \quad
w_{s2}^{12} = w_{s2}^{32} \, , \quad
w_{s2}^{21} = w_{s2}^{23} \, .
\label{eq: weights - relations}
\end{equation}
\label{eq: weights}
\end{subequations}

In Eqs.~\eqref{eq: structure functions explicit - 0phiK}-\eqref{eq: structure functions explicit - 4phiK} we have included the dependence on the final lepton masses. Although they can be dropped for $e^+ e^-$ production at sufficiently high energies (as in RHIC), we point out that the same structure also holds for heavier lepton and open heavy-quark pair (upon replacement of $M_\ell \to M_{Q}$) production, for which masses cannot be neglected. 

To verify the validity of our cross-section expressions we have compared them to results given in the literature. We generally find agreement up to mass terms that we keep. In Appendix~\ref{sec: cross check} the comparison with expressions in Refs.~\cite{Vidovic:1992ik, Klein:2020jom, Li:2019sin} is discussed for $M_\ell \ll M_\sT$, where a minus sign difference is found with respect to Ref.~\cite{Vidovic:1992ik}. In addition, we have verified that Eq.~\eqref{eq: structure functions explicit - 0phiK} agrees with Ref.~\cite{Pisano:2013cya} in the forward limit ($\Delta \to 0$), in which the photon GTMDs in Eq.~\eqref{eq: GTMDs} are related to photon TMDs.\footnote{Although the forward limit is not accessible in UPCs, it could be accessed in inclusive lepton pair production \cite{Pisano:2013cya}. However, it will require event selection and kinematic cuts to reduce possible QCD backgrounds.}

We note specifically the $\sin\phi$ dependences that are present in Eq.~\eqref{eq: cs dileqepton DeltaT-space}, which are absent in the forward (TMD) limit. Although nonzero in the off-forward case, one cannot select them using the corresponding sine-weight. This may seem counterintuitive, but is caused by the dependences on the azimuthal angles in the weights given in Eq.~\eqref{eq: weights}. Hence, to single out a structure function that involves a sine, a cosine (or sum of cosines) weight is instead required. Such operation may not isolate a single structure function however, but rather a combination of them due to the dependence of the GTMDs on $\bm k_{\sT}\cdot\bm \Delta_\sT$, which in the absence of gauge links and for zero skewness will only involve even powers\footnote{This is in contrast to the case of dihadron fragmentation functions discussed in Ref.~\cite{Boer:2003ya}, where $\sin\phi$ terms appear in the cross section due to dependence of the functions on odd powers of $\bm k_{\sT}\cdot\bm R_\sT$. In that case $\bm R_\sT$ is the difference in transverse momentum of the two final state hadrons (not to be confused with the transverse center of longitudinal momentum that we introduced previously). The odd powers do not require the inclusion of gauge links in that case.} of $\bm k_{\sT}\cdot\bm \Delta_\sT$. 
Therefore, in this paper we will assume that the GTMDs only depend on even powers of $\bm k_{\sT}\cdot\bm \Delta_\sT$, namely 
\begin{equation}
   {\cal F}_i^\gamma (x,\bm k_{\sT}^2, \bm \Delta_\sT^2, \bm k_\sT \cdot \bm \Delta_\sT) = \sum_{m = 0}^\infty F_i^{\gamma\, (2m)}\, \cos^{2m}\phi_{k\Delta}\, .
\label{eq: GTMD expansion}
\end{equation}
If anisotropies are not suppressed, all orders in $m$ need to be kept, leading to a feed-in mechanism of harmonics in $\phi_\Delta$ of a certain order into harmonics in $\phi_b$ of different orders. This we will discuss in detail in the next section, where we will explore the consequence of Eq.~\eqref{eq: GTMD expansion}.

\subsection{Expressions in \texorpdfstring{$b_\sT$}{b{\textunderscore}T}-space}\label{subsec b space}
In this section we present the dependence of the differential cross section on the impact parameter. To obtain it, we need to perform the Fourier transformation in Eq.~\eqref{eq:csFT}, which is a non-trivial operation because of the dependence of the GTMDs on $\bm k_{\sT}\cdot\bm \Delta_\sT$.
Schematically, if one assumes there is no $\bm k_{\sT}\cdot\bm \Delta_\sT$ dependence, or equivalently no $\bm{q}_\sT \cdot \bm{\Delta}_\sT$ dependence in the convolutions, one needs to consider integrals of the form:
\begin{align}
\int \frac{\d^2\bm \Delta_\sT}{(2\pi)^2} \, \e^{i \bm b_\sT \cdot \bm \Delta_\sT}\, \cos(2n \phi_{\Delta a})
\, {\cal C}(\bm  q_\sT^2, \bm \Delta_\sT^2) & = \int_0^\infty \frac{\d\Delta_\sT^2}{8\pi^2} \, {\cal C}(\bm  q_\sT^2, \bm \Delta_\sT^2) \, \int_0^{2\pi} \d\phi_\Delta \, \e^{i b_\sT \Delta_\sT \cos \phi_{\Delta b}}\, \cos(2n \phi_{\Delta a}) \nonumber \\
& = \frac{(-1)^n}{4\pi} \, \cos(2n \phi_{b a}) \int_0^\infty \d\bm \Delta_\sT^2\, J_{2n}(b_\sT\,\Delta_\sT)\, {\cal C}(\bm  q_\sT^2, \bm \Delta_\sT^2)\, , 
\end{align}
where $\phi_a$ is an unspecified azimuthal angle that could be $\phi_q$ or $\phi_K$ or a combination of them. The same conclusion also applies to the 
terms involving $\sin(2n\phi_{\Delta a})$. Thus, under this assumption, any isolated $\phi_\Delta$ dependence in $\d\sigma/\d^2\Delta_\sT$ is translated into an analogous $\phi_b$ dependent term 
in $\d\sigma/\d^2 b_\sT$ by employing the following one-to-one relations:
\begin{equation}
\begin{aligned}
    \int_0^{2\pi} \d \phi_\Delta \, \e^{i\, \bm b_\sT \cdot \bm \Delta_\sT} \cos(2n\phi_\Delta) & = (-1)^n 2\pi\, J_{2n}(b_\sT\,\Delta_\sT)\, \cos(2n\phi_b) \,, \\
    \int_0^{2\pi} \d \phi_\Delta \, \e^{i\, \bm b_\sT \cdot \bm \Delta_\sT} \sin(2n\phi_\Delta) & = (-1)^n 2\pi\, J_{2n}(b_\sT\,\Delta_\sT)\, \sin(2n\phi_b) \,.
\end{aligned}
\label{eq: cos&sin FT}
\end{equation}
On the other hand, if there is a $\bm{q}_\sT \cdot \bm{\Delta}_\sT$ dependence in the convolutions due to the internal angular dependence of the GTMDs, this one-to-one correspondence between $\cos(2n \phi_{\Delta a})$ and $\cos(2n \phi_{b a})$ terms is no longer present and an arbitrary number of trigonometric functions will appear: 
\begin{align}
\int \frac{\d^2\bm \Delta_\sT}{(2\pi)^2} \, \e^{i \bm b_\sT \cdot \bm \Delta_\sT}\, \cos(2n\phi_{\Delta a})\,  {\cal C}(\bm  q_\sT^2, \bm \Delta_\sT^2, \bm{q}_\sT \cdot \bm{\Delta}_\sT) & = \sum_{m = 0}^\infty \int_0^\infty \frac{\d\Delta_\sT^2}{8\pi^2} \, {\cal C}_m(\bm  q_\sT^2, \bm \Delta_\sT^2) \nonumber\\
& \phantom{=} \quad \times \int_0^{2\pi} \d\phi_\Delta \, \e^{i b_\sT \Delta_\sT \cos\phi_{\Delta b}}\, \cos(2n\phi_{\Delta a}) \, \cos^{2m} \phi_{\Delta q}\, ,
\end{align}
in analogy to the expansion in Eq.~\eqref{eq: GTMD expansion}.
For example, for $n=m=1$ and $\phi_a=2\phi_q$ the integral in the last expression yields:
\begin{align}
\int_0^{2\pi} \d\phi_\Delta \, \e^{i b_\sT \Delta_\sT \cos\phi_{\Delta b}}\, \cos2\phi_{\Delta q} \, \cos^{2} \phi_{\Delta q} = \pi \left(\frac{1}{2}\, J_0(b_\sT\,\Delta_\sT) - J_2(b_\sT\,\Delta_\sT) \cos2\phi_{b q} + \frac{1}{2}\, J_4(b_\sT\,\Delta_\sT) \cos 4\phi_{b q} \right)\,. 
\label{eq: feed-in example}
\end{align}
Thus, any modulation in $2n\phi_\Delta$ will “feed into” $2n' \phi_b$ modulations where $n'$ can be smaller, equal or larger than $n$, including the isotropic one ($n^\prime = 0$). One might expect such feed-in to be strongly suppressed, because anisotropies seem to be small for small-$x$ gluons in a nucleus \cite{Dumitru:2014dra,Mantysaari:2019csc,Caucal:2023fsf}, however, as we will see this is not the case for the photon GTMD.
If not suppressed, then all orders in $m$ will have to be taken into account. 
In principle this feed-in, or the suppression of it, could be tested in experiments, and the available data already supports the unsuppressed feed-in as will be discussed in section \ref{sec: model and numerical result}. 

For clarity we will first consider the terms in $b_\sT$-space that do have a one-to-one correspondence in $\Delta_\sT$-space. This means we first only consider $m=0$, such that ${\cal F}_i^\gamma = F_i^{\gamma\, (0)}$ and for which the differential cross section in $b_\sT$-space will read
\begin{align}
    \frac{\d\sigma}{\d {\rm PS}\,\d^2\bm b_\sT}
    \approx \frac{\alpha^2}{s_{NN} M_\sT^2} \int \frac{\d \bm \Delta_\sT^2}{4 \pi} & \bigg\{ J_0(b_\sT\, \Delta_\sT) \Big[
    F^0 + F^{\cos 2\phi_{qK}}\cos 2\phi_{qK}  + F^{\cos 4\phi_{qK}} \cos 4\phi_{qK} \Big]\nonumber\\
    & \phantom{=}- J_2(b_\sT\, \Delta_\sT) \Big[ F^{\cos 2\phi_{\Delta K}} \cos2\phi_{b K} + F^{\cos 2\phi_{q\Delta}} \cos 2\phi_{qb}\nonumber\\
    & \phantom{=} + F^{\cos 2 (\phi_{qK} + \phi_{\Delta K}) } \, \cos 2 (\phi_{qK} + \phi_{b K}) \Big] + J_4(b_\sT\, \Delta_\sT) \, F^{\cos 4 \phi_{\Delta K}} \cos 4\phi_{bK} \bigg\}\,.
\label{eq: cs dileqepton b space}
\end{align}
Note that, in contrast with Eq.~\eqref{eq: cs dileqepton DeltaT-space},  Eq.~\eqref{eq: cs dileqepton b space} does not contain any $\sin\phi$ term. It turns out that, in absence of feed-in, the integral over the azimuthal angles of the partonic transverse momenta, $k_{1\sT}$ and $k_{2\sT}$, removes all sines in $b_\sT$-space. The sines remain absent also when feed-in contributions are included, as we will explain at the end of this section.

We first compare Eq.~\eqref{eq: cs dileqepton b space}
 with the result obtained in Ref.~\cite{Shi:2024gex}, which is the only other paper thus far that considers the cross section differential in all three transverse vectors $\bm K_\sT$, $\bm q_\sT$ and $\bm b_\sT$. Although that paper only considers the GTMDs ${\cal F}^\gamma_1$ and ${\cal F}^\gamma_2$ in the parameterization of the GTMD correlator instead of all four, the employed model for the GTMD correlator does lead to nonzero ${\cal F}^\gamma_3$ and ${\cal F}^\gamma_4$, even if not identified explicitly. Ref.~\cite{Shi:2024gex} does not list the $\cos 2 \phi_{q K}$ and $\cos 2 \phi_{b K}$ asymmetries, but has an additional $\cos 4 \phi_{q b}$ term. As can be seen from Eq.~(\ref{eq: structure functions explicit - 2phiK}), the $\cos 2 \phi_{qK}$ and $\cos 2 \phi_{b K}$ terms are proportional to the square of the produced particle mass and therefore are negligible for electrons, but not necessarily for muons, charm and bottom quarks, hence we keep them. From Eq.~\eqref{eq: structure functions explicit} one sees that when ${\cal F}^\gamma_3$ and ${\cal F}^\gamma_4$ are dropped the $\cos 2 \phi_{b K}$, $\cos 4 \phi_{b K}$, $\cos 2 \phi_{q b}$ and $\cos 2 (\phi_{q K} + \phi_{b K})$ asymmetries should not be present, but using the adopted model which leads to nonzero ${\cal F}^\gamma_3$ and ${\cal F}^\gamma_4$, such asymmetries do appear. Concerning the $\cos 4\phi_{q b}$ modulation of Ref.~\cite{Shi:2024gex}, it does not appear in Eq.~\eqref{eq: cs dileqepton b space} as it comes from $(q_\sT \cdot \Delta_\sT)^2$ terms. 
If one includes such terms, further asymmetries would be generated as well, and indeed, in Ref.~\cite{Shi:2024gex} higher harmonics like $\cos (6\phi_{q K} - 2\phi_{b K})$ and $\cos({6\phi_{b K} - 2\phi_{q K}})$ are considered. These asymmetries arise from the feed-in mechanism we discussed above, but ever higher harmonics will be generated as well, of course. Other previous works have also studied certain specific asymmetries in this process, like the $\cos 2 \phi_{q K}$ and $\cos 4 \phi_{q K}$ in Ref.~\cite{Li:2019sin} and the $\cos 4\phi_{b K}$ in Ref.~\cite{Xiao:2020ddm}. 

If one includes next order terms ($m\geq 1$) in Eq.~\eqref{eq: GTMD expansion}, the expression in Eq.~\eqref{eq: cs dileqepton b space} will obtain additional \textit{cosine} asymmetries due to the feed-in, see e.g.~Eq.~\eqref{eq: feed-in example}.
Note that the feed-in does not affect $\phi_K$ and, consequently, we can organize the cross section into three groups. They are identified by the number of $\phi_K$ factors involved, with the feed-in causing the interference among the different structure functions within the same group. Moreover, as we have emphasized above Eq.~\eqref{eq: GTMD expansion} one can only observe cosine asymmetries in $b_\sT$-space, because the $\sin\phi$ terms present in Eq.~\eqref{eq: cs dileqepton DeltaT-space} only contribute via feed-in to cosine asymmetries in $b_\sT$-space (recall that the angle $\phi_\Delta$ is not directly observable in UPCs, while $\phi_b$ is).
To illustrate this, we consider as an example the distribution of the cross section with respect to the azimuthal angle difference $\phi_{qK}$:
\begin{align}
    \frac{\d\sigma}{\d\phi_{qK}} \approx \frac{\alpha^2}{s_{NN} M_\sT^2} \left( \widetilde F^0 + \widetilde F^{\cos4 \phi_{qK}} \cos 4\phi_{qK} + \widetilde F^{\sin 4\phi_{qK}} \sin 4\phi_{qK} \right)\, ,
\label{eq: cos4phi dependence}
\end{align}
where the approximation comes from neglecting the lepton masses. A priori, one might conclude from Eq.~\eqref{eq: cos4phi dependence} that a $\sin4\phi_{qK}$ term should be present alongside $\cos 4\phi_{qK}$. However, it turns out that $\widetilde F^{\sin 4\phi_{qK}}$ is zero upon integration over $\phi_{q}$.
The structure functions $\widetilde F$ (in which the feed-in contributions are hidden) are explicitly given by
\begin{equation}
\begin{aligned}
    \widetilde F^0 & = \int \frac{\d^2 \bm \Delta_\sT}{8\pi}\, \frac{\d^2 \bm b_\sT}{2\pi}\,\d \overline {\rm PS}\, \e^{-i \bm b_\sT \cdot \bm \Delta_\sT}\left[ F^0 + F^{\cos2\phi_{q\Delta}} \cos 2\phi_{q\Delta} + F^{\sin 2{\phi_{q\Delta}}} \sin 2\phi_{q\Delta} \right] \, , \\
    \widetilde F^{\cos4\phi_{qK}} & = \int \frac{\d^2 \bm \Delta_\sT}{8\pi}\, \frac{\d^2 \bm b_\sT}{2\pi}\, \d \overline {\rm PS}\, \e^{-i \bm b_\sT \cdot \bm \Delta_\sT}\, \Big[ F^{\cos4\phi_{qK}} + F^{\cos4\phi_{q\Delta}} \cos4\phi_{q\Delta} + F^{\cos2(\phi_{qK} + \phi_{\Delta K})} \cos2\phi_{q\Delta} \\
    & \hspace{5.1cm} - F^{\sin2(\phi_{qK} + \phi_{\Delta K})} \sin2\phi_{q\Delta} \Big]\, , \\
    \widetilde F^{\sin4\phi_{qK}} & = \int \frac{\d^2 \bm \Delta_\sT}{8\pi}\, \frac{\d^2 \bm b_\sT}{2\pi}\, \d \overline {\rm PS}\, \e^{-i \bm b_\sT \cdot \bm \Delta_\sT}\, \Big[ F^{\sin4\phi_{qK}} + F^{\cos4\phi_{q\Delta}} \sin4\phi_{q\Delta} + F^{\cos2(\phi_{qK} + \phi_{\Delta K})} \sin2\phi_{q\Delta} \\
    & \hspace{5.1cm} + F^{\sin2(\phi_{qK} + \phi_{\Delta K})} \cos2\phi_{q\Delta} \Big]\, ,
\end{aligned}
\end{equation}
where $\d \overline {\rm PS} = \d y_1\, \d y_2\, \d K_\sT^2\, \d q_\sT^2\, \d \phi_q$.
All terms on the right-hand side of $\widetilde F^{\sin4\phi_{qK}}$ involve an odd number of sine functions. Therefore, they are zero upon integration over $\phi_q$, namely
\begin{equation}
    \int_0^{2\pi} \d \phi_q\, \Big[ F^{\sin4\phi_{qK}} + F^{\cos4\phi_{q\Delta}} \sin4\phi_{q\Delta} + F^{\cos2(\phi_{qK} + \phi_{\Delta K})} \sin2\phi_{q\Delta} + F^{\sin2(\phi_{qK} + \phi_{\Delta K})} \cos2\phi_{q\Delta} \Big] = 0\ ,
    \label{eq:int-0}
\end{equation}
where we recall that the structure functions include convolutions and hence integrals over the azimuthal angles of internal transverse momenta. The result in Eq.~\eqref{eq:int-0} implies that $\widetilde F^{\sin4\phi_{qK}}$ is zero, independently of the integration over $\bm \Delta_\sT$ and $\bm b_\sT$ (including their azimuthal angles).
On the other hand, $\widetilde F^0$ and $\widetilde F^{\cos4\phi_{qK}}$ contain an even number of sine functions and are therefore nonzero upon the same integration.
The same argument applies to the other $\sin\phi$ asymmetries, leaving the $\cos\phi$ modulations as the only observable ones.

\section{Model studies and predictions}
\label{sec: model and numerical result}

Based on the results of Refs.~\cite{Li:2019sin, Vidovic:1992ik}, obtained within the EPA, we can express the correlator for the photon GTMD of an unpolarized nucleus as follows,\footnote{To see the correspondence with Eq.~(38) of~Ref.~\cite{Vidovic:1992ik}, one can take $\bm k_\perp \to \bm \kappa_\sT$ and  $\bm k_\perp \to \bm \kappa_\sT^\prime$ in the complex conjugate, and keep $\bm \kappa_\sT$ and $\bm \kappa_\sT^\prime$ as independent variables, to rewrite Eq.~(38) of~Ref.~\cite{Vidovic:1992ik} as 
$$f(x, \bm b_\sT) = (2\pi)^2 \int \frac{\d^2 \bm \kappa_\sT}{(2 \pi)^2} \int \frac{\d^2 \bm \kappa^\prime_\sT}{(2 \pi)^2}\, f_{\gamma/A} (x, \bm k_\sT, \bm \Delta_\sT)\, \e^{- i \bm b_\sT \cdot (\bm \kappa_\sT - \bm \kappa^\prime_\sT)}\, .$$}
\begin{equation}
    G_\gamma^{\mu\nu}(x, \bm k_\sT, 0, \bm \Delta_\sT) = \frac{\kappa_\sT^\mu\, {\kappa'}_{\!\sT}^{\nu}}{\bm \kappa_\sT \cdot \bm \kappa_\sT^\prime}\,  \frac{f_{\gamma/A} (x, \bm k_\sT, \bm \Delta_\sT)}{x}\, ,
\label{eq: corr EPA}
\end{equation}
with $\kappa_\sT^\mu = k_\sT^\mu - \frac{\Delta_\sT^\mu}{2}$ and ${\kappa'}_{\!\sT}^\mu = k_\sT^\mu + \frac{\Delta_\sT^\mu}{2}$.
Noting that 
\begin{align}
\kappa_\sT^\mu\, {\kappa'}_{\!\sT}^\nu = 
-\frac{1}{2}\left(\bm{k}_\sT^2 - \frac{1}{4} \bm{\Delta}_\sT^2\right) g_\sT^{\mu\nu} + k_\sT^{\mu\nu} - \frac{1}{4}\Delta_\sT^{\mu\nu}+
\frac{1}{2}k_\sT^{[\mu}\Delta_\sT^{\nu]},
\end{align}
and comparing to Eq.\ (\ref{eq: GTMDs}), one finds within this approximation
simple relations among the GTMDs ${\cal F}_i^\gamma$ through the expressions
\begin{equation}
\begin{aligned}
    {\cal F}_1^\gamma (x, \bm k_\sT, \bm \Delta_\sT) & = f_{\gamma/A} (x, \bm k_\sT, \bm \Delta_\sT)\, ,\\
    {\cal F}_2^\gamma (x, \bm k_\sT, \bm \Delta_\sT) & = \frac{2 M_N^2}{\bm \kappa_\sT \cdot \bm \kappa^\prime_\sT}\, f_{\gamma/A} (x, \bm k_\sT, \bm \Delta_\sT)\, ,\\
    {\cal F}_3^\gamma (x, \bm k_\sT, \bm \Delta_\sT) & = - \frac{1}{2}\frac{M_N^2}{\bm \kappa_\sT \cdot \bm \kappa^\prime_\sT}\, f_{\gamma/A} (x, \bm k_\sT, \bm \Delta_\sT)\, ,\\
    {\cal F}_4^\gamma (x, \bm k_\sT, \bm \Delta_\sT) & = \frac{M_N^2}{\bm \kappa_\sT \cdot \bm \kappa^\prime_\sT}\, f_{\gamma/A} (x, \bm k_\sT, \bm \Delta_\sT)\, ,
\end{aligned}
\label{eq: GTMDs EPA}
\end{equation}
implying
\begin{align}
{\cal F}_1^\gamma = \frac{\bm{k}_\sT^2 - \frac{1}{4} \bm{\Delta}_\sT^2}{2M_N^2}{\cal F}_{2}^\gamma = -2 \frac{\bm{k}_\sT^2 - \frac{1}{4} \bm{\Delta}_\sT^2}{M_N^2}{\cal F}_{3}^\gamma = \frac{\bm{k}_\sT^2 - \frac{1}{4} \bm{\Delta}_\sT^2}{M_N^2}{\cal F}_{4}^\gamma\,,  
\end{align}
in analogy to what has been found for gluons (the dipole gluon GTMDs to be specific) in the small-$x$ limit~\cite{Boer:2018vdi}.
The correlator in Eq.~\eqref{eq: corr EPA} is in agreement with the one used in Ref.~\cite{Shi:2024gex}. The sign of $\bm \Delta_\sT$ is not relevant since reversing it causes only a change of sign for ${\cal F}_4^\gamma$, without consequence on the cross section level since ${\cal F}_4^\gamma$ exclusively appears “squared”, as it is the only one that selects contributions antisymmetric in $\bm \Delta_\sT$ and $\bm k_\sT$.

The model in Eq.~\eqref{eq: corr EPA} incorporates unsuppressed $\bm k_\sT \cdot \bm \Delta_\sT$ dependencies to all orders, which will necessarily generate higher harmonics due to the feed-in we discussed before.
In order to assess the importance of this feed-in, we will introduce a parameter $\varepsilon$ in the GTMD model that can be used to suppress the $\bm k_\sT \cdot \bm \Delta_\sT$ dependence.
Specifically, following the relation between the photon distribution and the nuclear form factor (FF) $F_A(\bm k^2)$~\cite{Klein:2020jom}, we take 
\begin{equation}
    x\, f_{\gamma/A} (x, \bm k_\sT, \bm \Delta_\sT) = \frac{Z^2 \alpha}{\pi^2}\, \frac{\bm \kappa_\sT \cdot \bm \kappa_\sT^\prime}{ M(x, \kappa_\sT, \varepsilon)\, M(x, \kappa^\prime_\sT, \varepsilon)}
    \,F_A \big(\bm \kappa_\sT^2 + x^2 M_N^2 \big)\, F_A^*\big({\bm \kappa_\sT^\prime}^2 + x^2 M_N^2 \big) \, ,
\label{eq: Wigner distribution - Form Factor}
\end{equation}
where $M(x, \kappa_\sT, \varepsilon)$ is a normalization factor defined as
\begin{equation}
    M(x, \kappa_\sT, \varepsilon) = \bm \kappa_\sT^2 + (1 - \varepsilon)\, \bm k_\sT \cdot \bm \Delta_\sT + x^2 M_N^2\, ,
\label{eq: GTMD normalization}
\end{equation}
with $|\varepsilon| \leq 1$ and where the GTMDs adopted in {Refs.}~\cite{Klein:2020jom, Shi:2024gex} are retrieved for $|\varepsilon| = 1$. 
We point out that for $\kappa_\sT^\prime$ the sign of $\bm \Delta_\sT$ is reversed.
While the product $\kappa_\sT \cdot \kappa_\sT^\prime$ does not depend on the scalar product $\bm k_\sT \cdot \bm \Delta_\sT$ for the symmetric choice that we made in Eq.~\eqref{eq: kappa definition}, the FF does.
However, in the model adopted in Eq.~\eqref{eq: Wigner distribution - Form Factor} most of the feed-in contributions come from the prefactor and not from the FF. Consequently, the parameter $\varepsilon$ correctly suppresses $\bm k_\sT \cdot \bm \Delta_\sT$ dependences, and regulates the strength of the feed-in; in particular, we have found that $\varepsilon = 0.5$ is already sufficient to strongly suppress it. Accordingly, in the following we will show numerical results employing two values of this parameter: $\varepsilon = 1$ and $\varepsilon = 0.5$.
We also point out that the definition in Eq.~\eqref{eq: Wigner distribution - Form Factor} forces the GTMD to zero if either $\bm \kappa_\sT$ or $\bm \kappa_\sT^\prime$ is zero, or if they are perpendicular to each other.

Although the FFs of nuclei are in general not very well-known, a reasonable assumption is to consider the charge of a nucleus to be distributed according to the Woods-Saxon potential~\cite{Woods:1954zz}, which leads to~\cite{Li:2019yzy} 
\begin{equation}
    F_A(\bm k) = \frac{4\pi \rho_0}{A} \int_0^\infty \d r\,  \frac{r^2}{1 + \exp\! \big[(r - R_A)/d\big]}\, \frac{\sin(|\bm k| r)}{|\bm k| r}\ ,
\label{eq: Woods-Saxon FF}
\end{equation}
where $\bm k$ is a two-dimensional vector that belongs to the transverse plane, $\rho_0$ is the density of the nucleus, $R_A$ its radius, while $d$ describes the diffuseness of the nuclear surface. Considering gold nuclei, we take $R_A \approx 6.4$~fm and $d = 0.535$~fm. 
A numerical approximation to the Woods-Saxon potential that is extensively adopted in the literature corresponds to the FF expression used by STARlight~\cite{Klein:2016yzr}, which reads
\begin{equation}
    F_A(\bm k) = \frac{4 \pi \rho_0}{A\, |\bm k|^3} \frac{1}{(a\, |\bm k|)^2 + 1} \left[ \sin(|\bm k|\, R_A) - |\bm k|\, R_A \cos(|\bm k|\, R_A) \right]\, ,
\label{eq: Starlight FF}
\end{equation}
with $a = 0.7$~fm.
Besides parameterizing the FF, one can also parameterize the photon GTMD directly, thus ignoring Eq.~\eqref{eq: Wigner distribution - Form Factor}. In particular, we propose the following parameterization
\begin{equation}
    x\, f_{\gamma/A}(x, \bm k_\sT, \bm \Delta_\sT) = \frac{Z^2 \alpha}{\pi^2}\,\frac{1 + (1 - x)^2}{2}
    \frac{\bm \kappa_\sT \cdot \bm \kappa_\sT^\prime}{M(x, \kappa_\sT, \varepsilon)\, M(x, \kappa^\prime_\sT, \varepsilon)}\,
    \exp\left[ - \frac{D(x)}{Q_0^2} \left( \bm k_\sT^2 + \frac{\bm \Delta_\sT^2}{4} \right) \right]\, ,
\label{eq: GTMD gaussian ansatz}
\end{equation}
where the normalization factor $M$ is given in Eq.~\eqref{eq: GTMD normalization} and we take $Q_0 = 0.06$~GeV. In addition $D(x)$ is a profile function that accommodates the possibility that the width of the Gaussian varies with the photon energy; for simplicity, we take $D(x) = 1$. 

\begin{figure}[t]
\centering
\subfloat[\label{fig: bT dependence}]{\includegraphics[width=.485\linewidth, keepaspectratio]{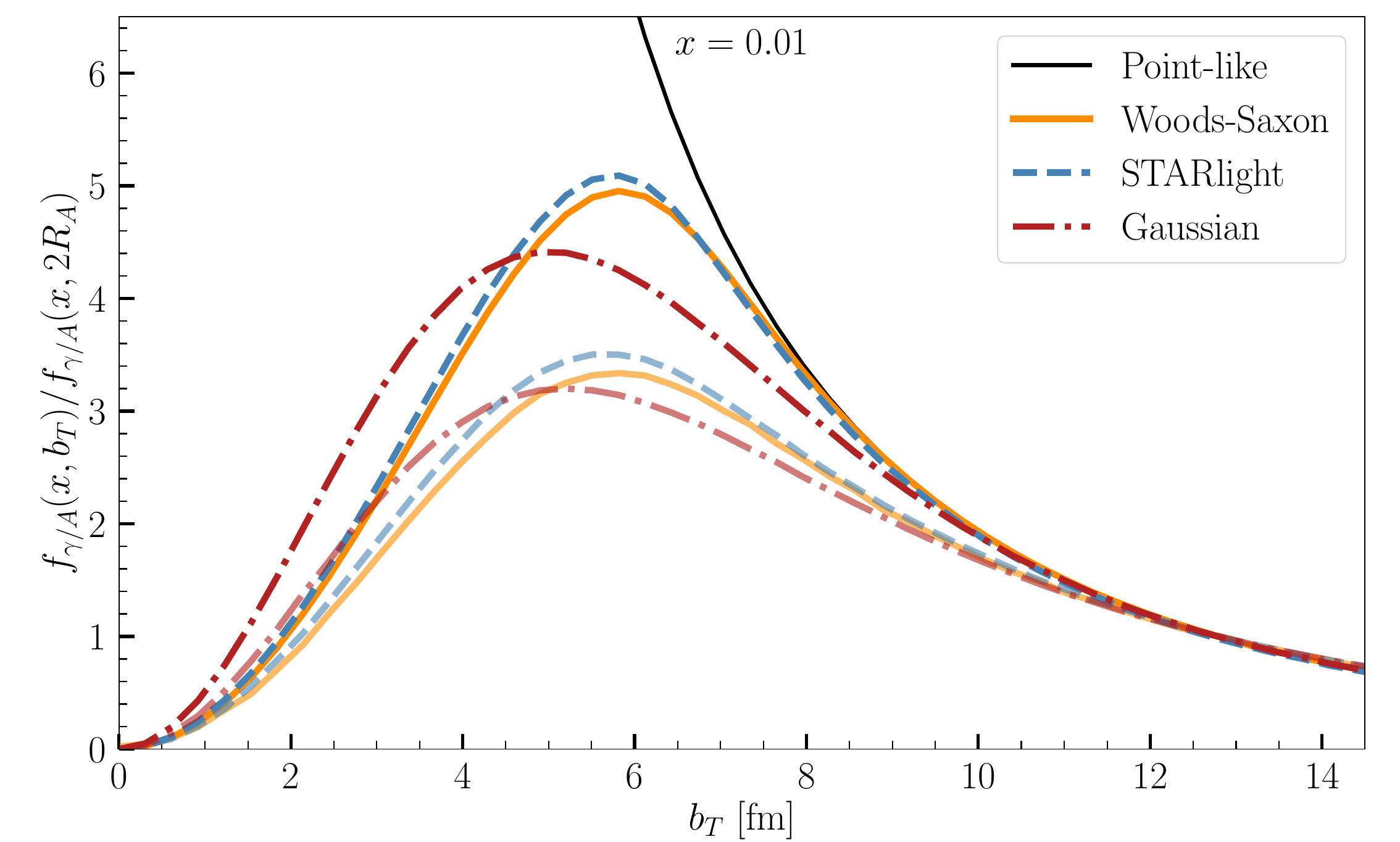}}\hfill
\subfloat[\label{fig: kT dependence}]{\includegraphics[width=.485\linewidth, keepaspectratio]{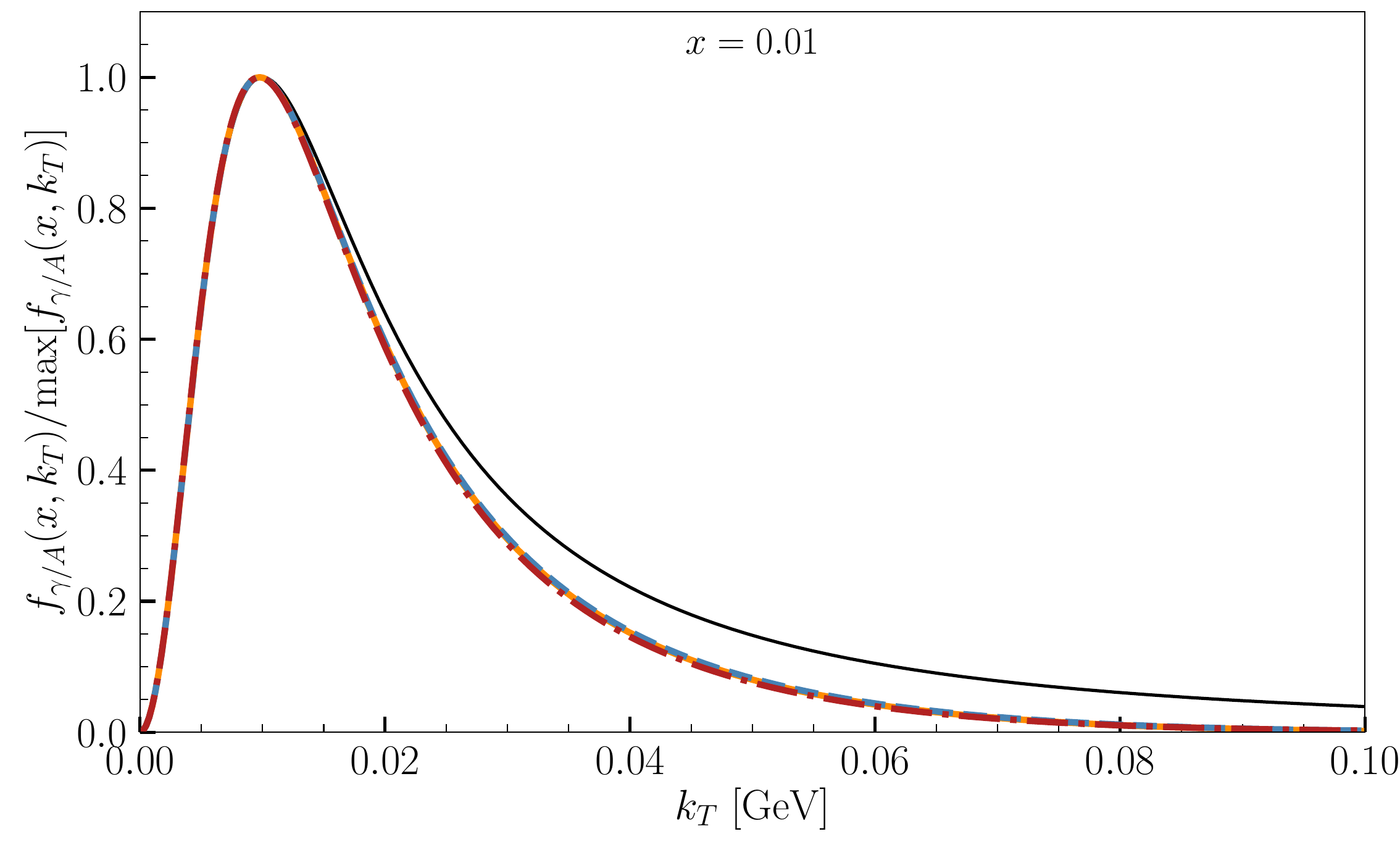}}
\caption{\it
(a) The impact parameter dependence of the $k_\sT$-integrated GTMD; (b) the $k_\sT$ dependence for $\Delta_\sT \to 0$. The orange solid line is obtained by employing the Woods-Saxon FF in Eq.~\eqref{eq: Woods-Saxon FF}, whereas the blue dashed line corresponds to the STARlight FF in Eq.~\eqref{eq: Starlight FF} with $a = 0.7$~fm. The red dash-dotted line refers to the model in Eq.~\eqref{eq: GTMD gaussian ansatz}. Thicker lines are obtained with $\varepsilon = 1$, and thinner ones with $\varepsilon = 0.5$.
The black solid thin line corresponds to the point-like curve (see Eq.~\eqref{eq: pointlike photon distribution}).}
\label{fig: GTMD dependences}
\end{figure}

Independently of the GTMD model employed, for sufficiently high  $b_\sT$ (at least greater than the radius of the nucleus), the photon distribution must become the same as that of a point-like source of charge $Z e$ in Eq.~\eqref{eq: pointlike photon distribution}, which is retrieved by taking $F_A^{\rm pl}(\bm k) \approx 1$ in Eq.~\eqref{eq: Wigner distribution - Form Factor}.\footnote{More precisely, we should have $|F_A^{\rm pl}(\bm k)|^2 = \frac{1 + (1 - x)^2}{2}$ to match exactly with Eq.~\eqref{eq: photon distribution EPA}, but for $x \ll 1$ this is approximately $1$.}
A comparison among the various GTMDs introduced above and the point-like distribution is given in Fig.~\ref{fig: GTMD dependences}. In panel (a) we show the $k_\sT$-integrated $b_\sT$-dependence of the distributions normalized by their value at $b_\sT = 2R_A$, whereas panel (b) presents the $k_\sT$-dependence in the forward limit of the same distributions normalized by their peak values; in both panels we have taken $x = 0.01$. Although the size of the distributions varies, their general behavior does not change between the $\varepsilon = 1$ and $\varepsilon = 0.5$ choices. 
All GTMD models considered here have the expected fall-off at high-$b_\sT$ values, which follows the point-like distribution. 
However, effects due to the nuclear structure can be observed also for $b_\sT > R_A$.
For gold nuclei, we have that the GTMD models differ from each other and deviate from the point-like ansatz for $b_\sT \lesssim 10~(12)~{\rm fm}$ for $\varepsilon = 1~(0.5)$.
In $k_\sT$-space, the GTMD models are in accordance with the point-like distribution only at low $k_\sT$, with all distributions peaking at the same value of $k_\sT$. Beyond this peak, all GTMD models deviate from the point-like distribution, yet staying in good agreement with each other. 
This deviation arises from the presence of the photon momentum fraction in Eqs.~\eqref{eq: Wigner distribution - Form Factor} and~\eqref{eq: GTMD gaussian ansatz}, which implies that for lower values of $x$ this difference greatly reduces.

Irrespective of the GTMD model employed, we can analytically integrate over the impact parameter $b_\sT$ when focusing exclusively on UPCs. In particular, for terms independent of $\phi_b$, which include the isotropic one~\cite{Klein:2020jom}, we have that
\begin{align}
    \frac{\d \sigma}{\d {\rm PS}} = \int \d^2 \bm b_\sT\, \frac{\d \sigma}{\d {\rm PS}\, \d^2 \bm b_\sT} 
    & = \int \d^2 \bm \Delta_\sT\,  \Big[ \delta^{(2)}(\bm \Delta_\sT) - \frac{b_{\min}}{2\pi \Delta_\sT}\, J_1(b_{\min}\, \Delta_\sT) \Big]\, \frac{\d \sigma}{\d {\rm PS}\, \d^2 \bm \Delta_\sT}\, ,
\label{eq: UPC b-integrated formula - 0}
\end{align}
with $b_{\min} = 2R_A$ and where the first term in brackets corresponds to the forward limit.
Besides, if we integrate the cross section with a $\phi_b$-dependent weight $W(\phi_{ba})$, we will have analogous formulae involving different Bessel functions. In more details, for $W(\phi_{ba}) = \cos 2 \phi_{ba}$ we obtain
\begin{align}
    \int \d^2 \bm b_\sT\, \frac{\d \sigma}{\d {\rm PS}\, \d^2 \bm b_\sT} \cos 2 \phi_{ba} = - \int \frac{\d^2 \bm \Delta_\sT}{2\pi \Delta_\sT} \left[b_{\min}\, J_1(b_{\min}\, \Delta_\sT) + \frac{2}{\Delta_\sT}\, J_0(b_{\min}\, \Delta_\sT) \right] \frac{\d \sigma}{\d {\rm PS}\, \d^2\bm \Delta_\sT}\, \cos 2\phi_{\Delta a}\, ,
\label{eq: UPC b-integrated formula - 2phi}
\end{align}
whereas for $W(\phi_{ba}) = \cos 4\phi_{ba}$
\begin{align}
    & \int \d^2 \bm b_\sT\, \frac{\d \sigma}{\d {\rm PS}\, \d^2 \bm b_\sT} \cos 4\phi_{ba} = \int \frac{\d^2 \bm \Delta_\sT}{2\pi \Delta_\sT} \left[ \frac{4}{\Delta_\sT}\, \Big( 3\, J_2(b_{\min}\, \Delta_\sT) + J_0(b_{\min}\, \Delta_\sT)\Big) - b_{\min}\, J_1(b_{\min}\, \Delta_\sT) \right] \frac{\d \sigma}{\d {\rm PS}\, \d^2\bm \Delta_\sT}\, \cos 4\phi_{\Delta a}\, .
\label{eq: UPC b-integrated formula - 4phi}
\end{align}
Note that Eqs.~\eqref{eq: UPC b-integrated formula - 2phi} and~\eqref{eq: UPC b-integrated formula - 4phi} involve asymmetries solely observable in the off-forward case, and are therefore unique to UPCs (or more precisely to non-central collisions). The same logic will also apply to modulations of higher orders in $\phi_{ba}$.
The formulae above can be exploited to isolate asymmetries in $b_\sT$ space that have no analogous term in $\Delta_\sT$ space, but that arise due to the anisotropy of the GTMDs (see e.g.\ Eq.~\eqref{eq: feed-in example}). As we will show later, a striking example is the $\cos 4\phi_{bq}$ asymmetry: a (sizable) $\cos 4\phi_{b q}$ term is generated by integrating the differential cross section in Eq.~\eqref{eq: cs dileqepton DeltaT-space} (that itself has no $\cos 4\phi_{\Delta q}$ term) with the weight $\cos 4\phi_{\Delta q}$.
In general, the above expressions show that for these weighted integrals there {\it is} a one-to-one correspondence between the weight in $b_\sT$ and that in $\Delta_\sT$ space.

\begin{figure}[t]
\begin{center}
\subfloat[\label{fig: dcs with diff GTMDs}]
{\includegraphics[width=.495\linewidth, keepaspectratio]{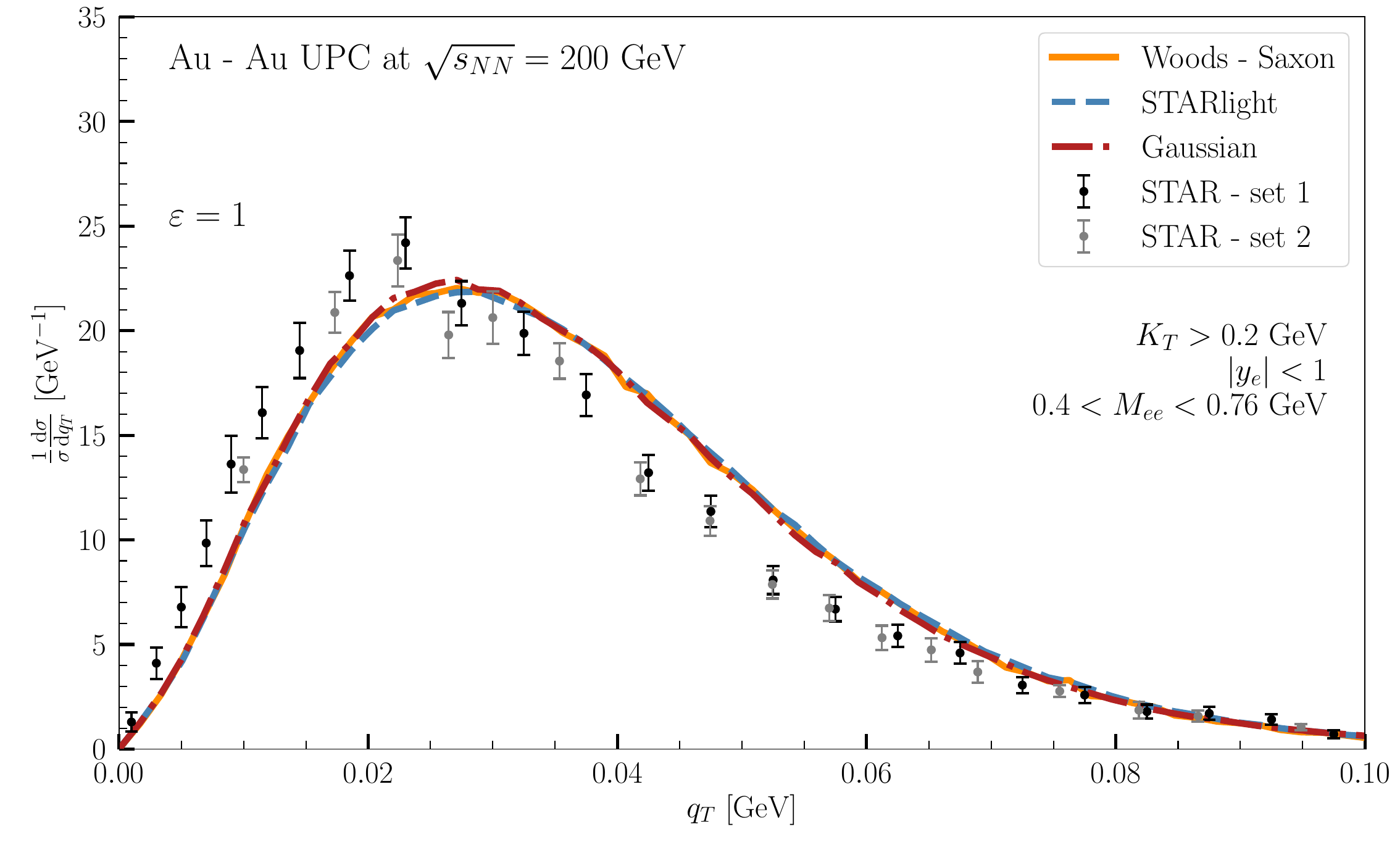}}\hfill
\subfloat[\label{fig: dcs contributions}]
{\includegraphics[width=.495\linewidth, keepaspectratio]{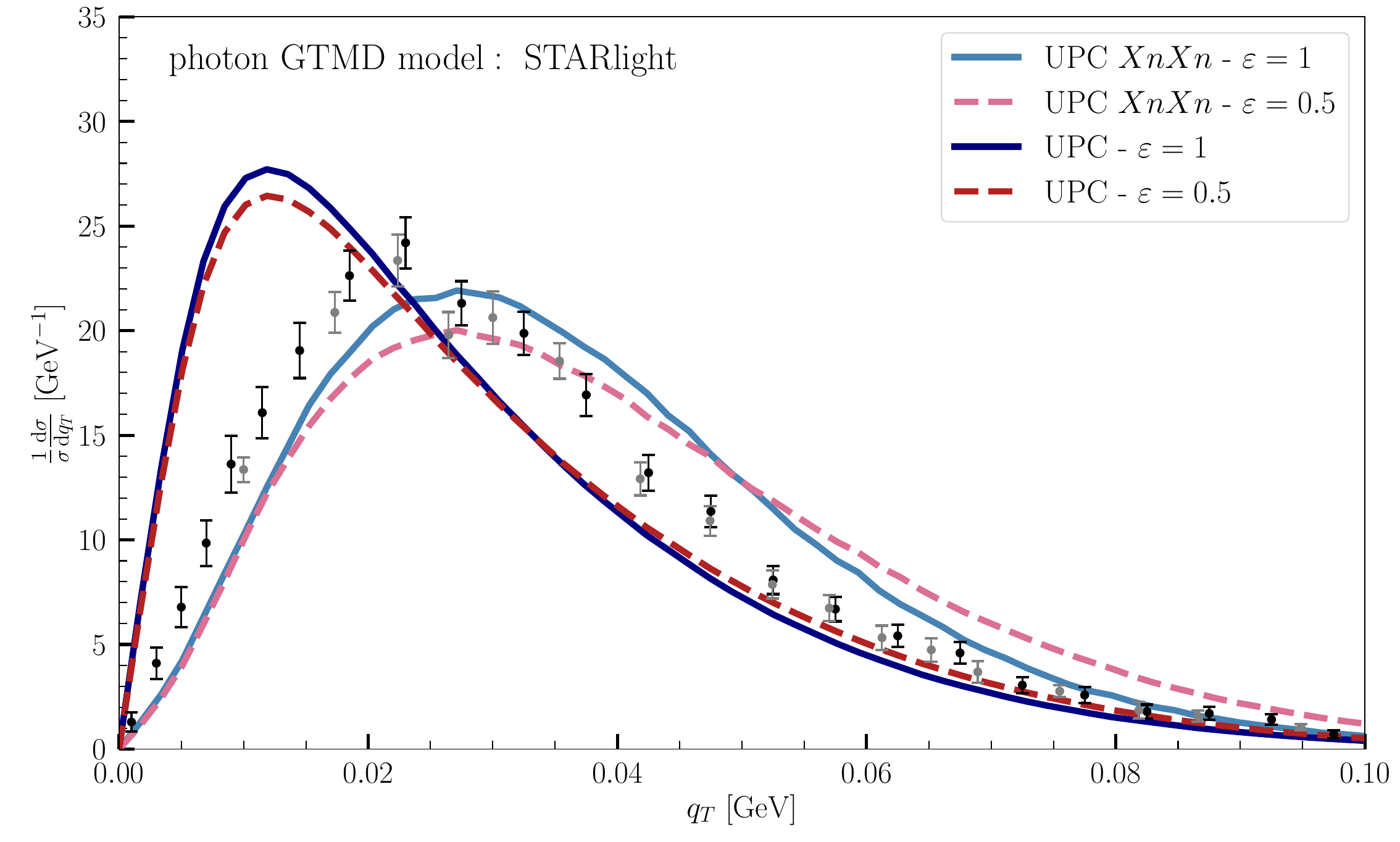}}
\caption{\it (a)  Normalized differential cross section as a function of $q_\sT$ for $e^+ e^-$ production in Au~-~Au UPC $Xn Xn$ at $\sqrt{s_{NN}} = 200\ {\rm GeV}$ for different models of the GTMDs and compared to STAR data~\cite{STAR:2019wlg}. Kinematical constraints are given in the legend. 
(b) Variation of the normalized cross section with respect to $\varepsilon$ for the full UPC and the $Xn Xn$ channel.
Curves are obtained using the STARlight FF and considering the same kinematical cuts used in (a). The STAR datasets are also included for comparison.
} 
\label{fig: STAR norm cross section}
\end{center}
\end{figure}

\begin{figure}[t]
\begin{center}
\includegraphics[width=.7\linewidth, keepaspectratio]{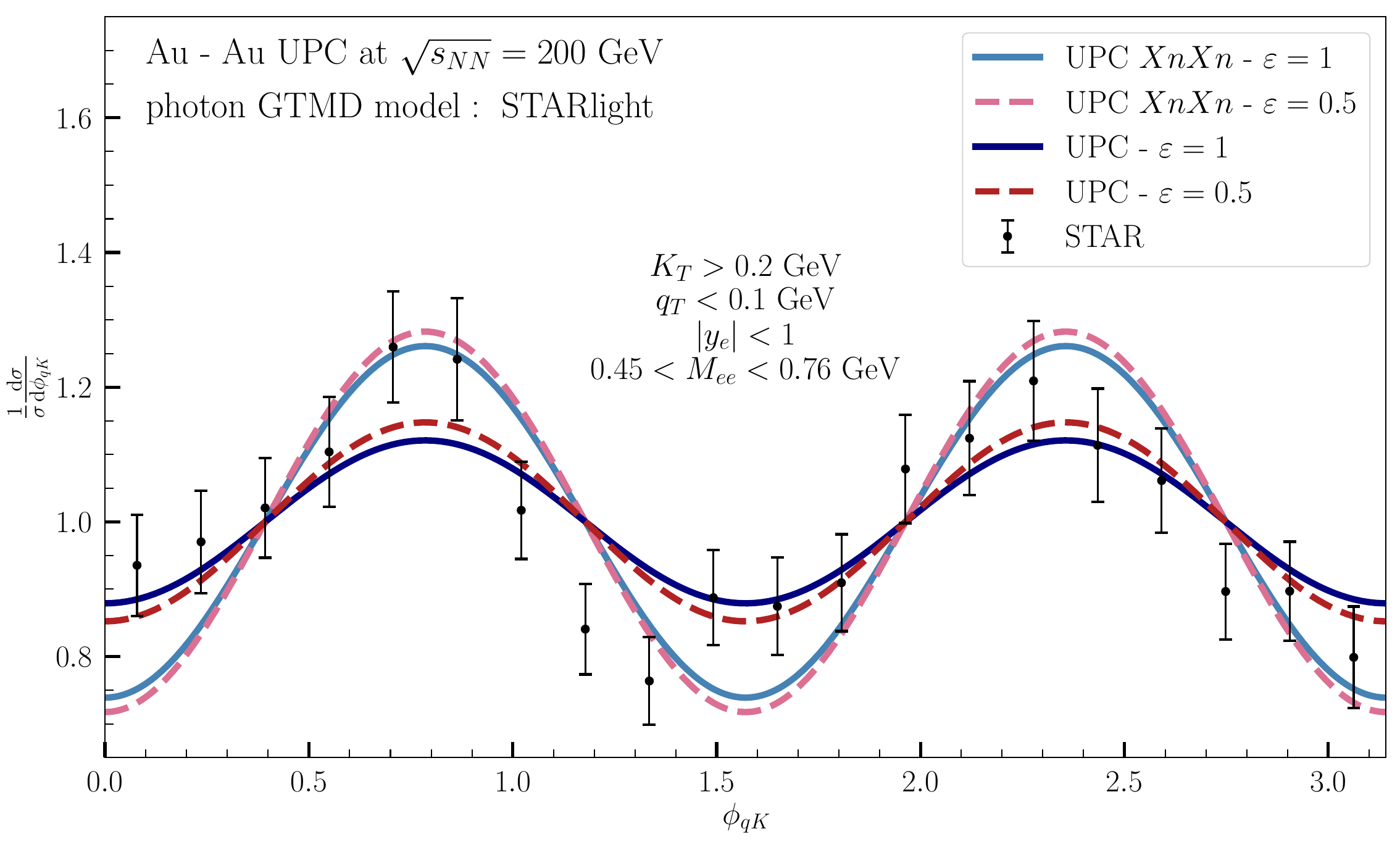}
\caption{\it Normalized cross section for $e^+ e^-$ production in Au~-~Au UPCs at $\sqrt{s_{NN}} = 200\ {\rm GeV}$ with respect to $\phi_{qK} = \phi_q - \phi_K$, for $\varepsilon = 1$ (solid blue lines) and $\varepsilon = 0.5$ (dashed pink lines).
The overall $\phi_{qK}$ modulation is compared with STAR data~\cite{STAR:2019wlg}.}
\label{fig: STAR cos4phi}
\end{center}
\end{figure}

To estimate the impact of the feed-in effect and test the accuracy of the GTMD models employed, in Figs.~\ref{fig: STAR norm cross section} and~\ref{fig: STAR cos4phi} we compare model predictions with STAR data~\cite{STAR:2019wlg} for the differential cross section with respect to $q_\sT$ and $\phi_{qK} \equiv \phi_q - \phi_K$, respectively. The experiment observed the production of $e^+e^-$ pairs in UPCs of two gold nuclei with a c.m.\ energy per nucleon-nucleon collision $\sqrt{s_{NN}} = 200~{\rm GeV}$. The leptons are produced at central rapidity ($-1 < y_e < 1$). Moreover, kinematical cuts are applied for the individual transverse momenta of the final leptons, such that ${K_\sT > 0.2~{\rm GeV}}$ and $q_\sT < 0.1$, which is in line with the correlation limit, and for the invariant mass of the lepton pair: ${40 < M_{ee} < 760~{\rm MeV}}$ for $\d\sigma/\d q_\sT$ and ${45 < M_{ee} < 760~{\rm MeV}}$ for $\d\sigma/\d \phi_{qK}$.
The STAR collaboration provides two datasets for the differential cross section with respect to $q_\sT$. For the first one (\textit{set $1$} in the following), it is specified that data are collected with one or more neutrons detected in the beam directions, whereas this information is lacking for the other one (\textit{set $2$}). 
However, since the two datasets are compatible with each other, we present in Fig.~\ref{fig: STAR norm cross section} both. To focus on the contribution of the GTMD tails rather than their overall normalization, we normalize the data with respect to the (discrete) integrated cross section, which is evaluated by summing the central points of experimental data weighted by the width of the corresponding $q_\sT$ bins. 
On the other hand, only one dataset is given for the $\d \sigma/\d \phi_{qK}$, for which the information regarding the number of emitted neutrons measured is again missing. From the discussion below it seems reasonable to assume that the unspecified datasets are also for the $Xn Xn$ case. 

To account for the number of neutrons emitted by each nucleus along the beam direction in the theoretical calculations, one has to weigh the differential cross section with a distribution $P_{NnNn}(b_\sT)$ that evaluates the probability of emitting a number $n$ of neutrons from the nucleus. The distribution depends on the impact parameter $b_\sT$, as one would expect fewer neutrons the more the two nuclei are separated. 
In particular, the probability of emitting at least one neutron per nucleus, which is based on the Poisson distribution (see also Ref.~\cite{Baltz:2002pp}), is given by~\cite{Shi:2024gex} (there called $P_{MnMn}$):
\begin{equation}
    P_{XnXn}(b_\sT) = \left( 1 - \e^{- P_s(b_\sT)} \right)^2 \, ,
\label{eq: probability 1+ neutron emission}
\end{equation}
where $P_s(b_\sT) = {94\ {\rm fm}^2}/b_\sT^2$ is the probability of emitting one neutron from a gold nucleus. Note that the normalization of $P_s$ is not unique, and we followed the prescription provided in Ref.~\cite{Baur:1998ay}.
With this, the differential cross section when at least one neutron per nucleus is emitted is given by
\begin{equation}
    \frac{\d \sigma}{ \d {\rm PS}} = \int_{b_{\min}}^\infty \d b_\sT\, b_\sT \int_0^{2\pi} \d \phi_b\, \frac{\d \sigma}{ \d {\rm PS}\, \d^2 \bm b_\sT}\, P_{XnXn}(b_\sT)\, .
\label{eq: cross section UPC - XnXn channel}
\end{equation}
Note that Eq.~\eqref{eq: UPC b-integrated formula - 0} is recovered upon inclusion of all the neutron emission channels.

From Fig.~\ref{fig: dcs with diff GTMDs}, we see that the predictions for the normalized cross section employing the three aforementioned GTMD models with $\varepsilon = 1$ are very comparable to each other and in reasonable but not full agreement with STAR data, but we consider this sufficiently close for our purposes.
We remark that the curves are obtained by means of Eq.~\eqref{eq: cross section UPC - XnXn channel} ($Xn Xn$ channel in UPCs) and without the inclusion of photons directly emitted by up and down quarks in the nucleons, a contribution that is negligible for UPCs.
In Fig.~\ref{fig: dcs contributions} we examine the impact of the parameter $\varepsilon$ by considering the full feed-in case of $\varepsilon=1$ and reducing it by a factor 2, which we have found to be already enough to strongly suppress feed-in contributions from the GTMDs. 
We will see that the former choice is closest to both STAR datasets overall. 
This suggests that feed-in contributions, which are generated from $F^{\cos 2\phi_{q\Delta}}$ and $F^{\sin 2 \phi_{q\Delta}}$, are not negligible. The former is the primary source of this effect, while the latter is required to ensure positive cross sections at all $b_\sT$ values. 
The difference between the two $\varepsilon$ choices becomes somewhat less pronounced for the full UPC result (i.e.\ the one including the $0n0n$ and $0n1n$ channels) that is more peaked.
The difference between the full UPC and $XnXn$ distributions, and in particular their peak positions, might suggest the possibility of using the latter to fix $P_s$. However, this involves two complications. 
The first one is the requirement of a very precise resolution in $q_\sT$, namely at the MeV order. Hence, we expect that experiments such as those at the LHC, which typically have a resolution of $50\textrm{-}100$~MeV, are not able to determine the peak position with sufficient precision.
The second issue regards the uncertainties related to $b_{\rm min}$. In fact, changes in $b_{\rm min}$ and $P_s$ cause comparable deviations of the peak position and distribution shape. Thus, this will hamper the extraction of $P_s$.

In Fig.~\ref{fig: STAR cos4phi} we show the dependence of the UPC cross section on the azimuthal angle difference $\phi_{qK}$. Since the various models yield highly similar results, from now on we will employ only the STARlight FF. 
The prediction curves for both $\varepsilon = 1$ and $\varepsilon = 0.5$ are compatible with STAR data, but their precision is insufficient to discriminate among them.
We point out that our result for $\varepsilon = 1$ with the inclusion of all neutron emission channels is slightly different from the theoretical curve reported in Ref.~\cite{STAR:2019wlg} which makes use of the expressions in Ref.~\cite{Li:2019sin}. We identified the source of such difference with the feed-in contribution from $F^{\sin2(\phi_{qK} + \phi_{\Delta K})}$ that has to be taken into account but does not seem to have been included in that reference.

\begin{figure}[t]
{\includegraphics[width=.7\linewidth, keepaspectratio]{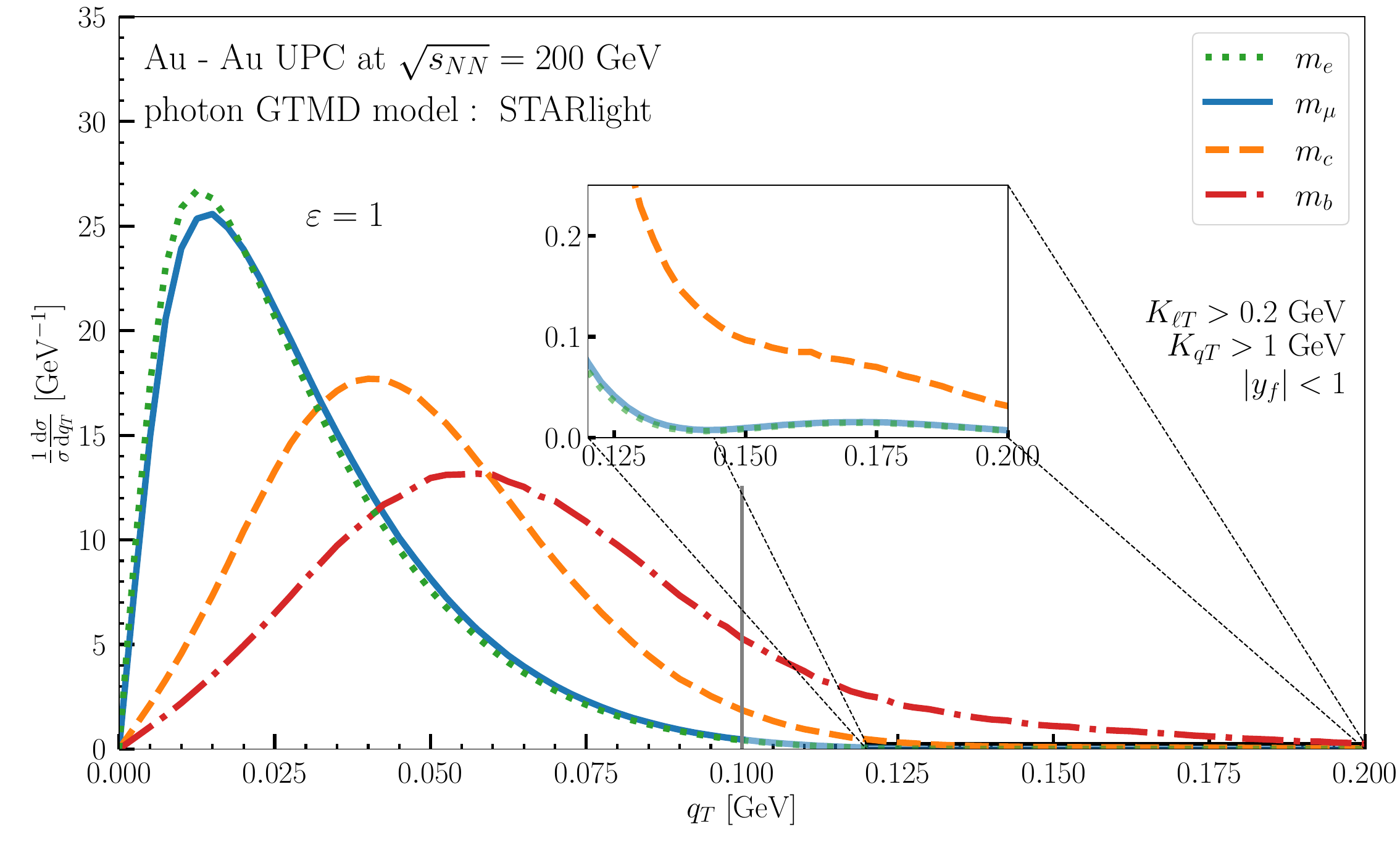}}
\caption{\it Mass dependence of the normalized differential cross section obtained using the STARlight FF. The green dotted line corresponds to $e^+ e^-$ production, blue solid to $\mu^+ \mu^-$, orange dashed to $c \bar c$, and the red dash-dotted one to $b \bar b$. For the leptons we integrate $K_{\ell \sT} > 0.2~{\rm GeV}$, whereas we take $K_{q \sT} > 1~{\rm GeV}$ for heavy quarks. The vertical line corresponds to the point after which we cannot trust the factorization for the lighter states. The zoom-in inset shows the tail of dilepton distributions.} 
\label{fig: mass dependence}
\end{figure}

In Fig.~\ref{fig: mass dependence} we explore the dependence of the normalized differential cross section on the masses of the final particles. We consider two dilepton systems, $e^+ e^-$ and $\mu^+ \mu^-$, and two heavy-quark pair productions, $c \bar c$ and $b \bar b$, in Au-Au UPCs at RHIC energies, namely $\sqrt{s_{NN}} = 200~{\rm GeV}$, and central rapidity of final particles, $|y_f| < 1$. We do not impose constraints on the invariant mass of the produced pair and from now on we only take $\varepsilon = 1$. We also consider a different $K_\sT$ integration between the dilepton systems ($K_\sT > 0.2~{\rm GeV}$) and the open heavy-quark production ($K_\sT > 1~{\rm GeV}$), since the reconstruction of $D^0$ mesons from kaon-pion pairs with a transverse momentum difference of $K_\sT = {\cal O}(0.1~{\rm GeV})$ between the two $D^0$-mesons might be too experimentally challenging (if they are also close in rapidity and azimuthal angles). Nevertheless, we have checked that including lower values of $K_\sT$ (down to $0.2$~{\rm GeV}) does not significantly modify the shape of the normalized cross section of heavy quark production.
Due to the higher masses, we present the cross sections of the open quark pair production up to $q_\sT = 0.2~{\rm GeV}$, after which they become negligible. 
On the other hand, for these choices of the kinematical cuts we cannot trust the factorization in the case of the lighter states beyond $q_\sT = 0.1~{\rm GeV}$, as indicated by the vertical gray line in the figure. Consequently, the cross sections for quarks and leptons are normalized taking $q_\sT$ up to $0.2$~GeV for the former and $0.1$~GeV for the latter.
We observe that the normalized cross section is not very sensitive to the difference between the muon and electron masses and the two distributions become indistinguishable from each other with higher cuts (e.g.~$K_\sT > 0.4~{\rm GeV}$). Therefore one could consider adding the $e^+ e^-$ and $\mu^+ \mu^-$ data to increase statistics, but as we will see this is not the case for the azimuthal modulations. 
For heavy-quark production the cross-section becomes more sensitive to the heavier masses and the distribution becomes wider, 
with its peak shifting towards higher $q_\sT$ values.
Another difference between the dilepton and heavy-quark pair cases is the role of the off-forward contribution in Eq.~\eqref{eq: UPC b-integrated formula - 0}. 
Indeed, while the UPC and forward limit distributions for dilepton productions are peaked at similar $q_\sT$, for heavy quarks the peak of the UPC distribution is shifted to lower $q_\sT$ compared to the forward limit.
A final remark concerns the tail of the dilepton distributions in $q_\sT$. The zoomed figure highlights the presence of an oscillating behavior that is dependent on the $b_{\min}$ in the $b_\sT$ integration.
 Therefore, the same also occurs in the production of heavier states, as can be noticed in the tail of the charm distribution. However, the oscillations are also enhanced by the fact that the dilepton distributions are more peaked; since the mass-dependent terms of the cross section cause the broadening of the distribution, the oscillations are smoothed out for heavy quark production. We will elaborate more on this effect below since this oscillatory behavior in some cases strongly affects the asymmetries as well. It should be pointed out that the inclusion of Sudakov resummation may suppress the oscillations and thus have a significant effect on the asymmetries too, as was discussed for di-muon production in Refs.~\cite{Shao:2022stc, Shao:2023zge}. Therefore, future UPC data may show the importance of such Sudakov effects through the absence or presence of the oscillations and related features in the various asymmetries that we will discuss below.

\begin{figure}[t]
\begin{center}
{\includegraphics[width=.7\linewidth, keepaspectratio]{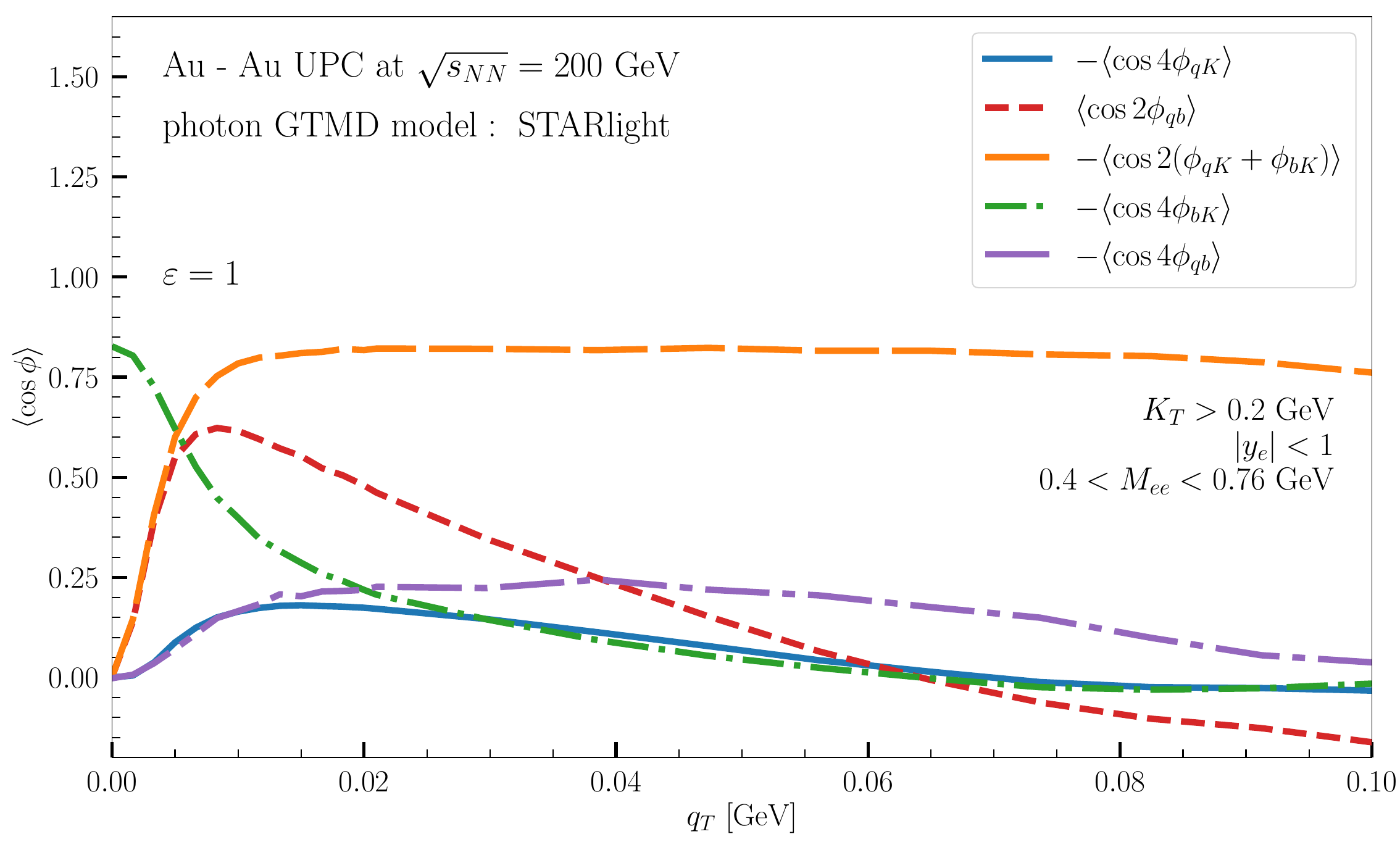}}
\caption{\it Asymmetries for $e^+ e^-$ production in Au~-~Au UPCs at RHIC.}
\label{fig: asymmetries ee}
\end{center}
\end{figure}

Now we turn to the azimuthal asymmetries expected in this process.
The normalized asymmetry in a certain azimuthal angle $\phi$ is defined as
\begin{equation}
    \langle \cos \phi \rangle = 2\, \frac{\int \d \phi\, \d {\rm PS}^\prime\ \d \sigma\, \cos\phi}{\int \d \phi\, \d {\rm PS}^\prime\ \d \sigma}\, ,
\end{equation}
where $\d\sigma \equiv \d \sigma/\d {\rm PS}\, \d^2 \bm b_\sT$, $\d {\rm PS}^\prime$ is a short-hand notation that indicates the presence of other integration variables, and we have multiply the fraction by a factor $2$ such that $|\langle \cos \phi \rangle | \leq 1$.

\begin{figure}[t!]
\begin{center}
\subfloat[]
{\includegraphics[width=.333\linewidth, keepaspectratio]{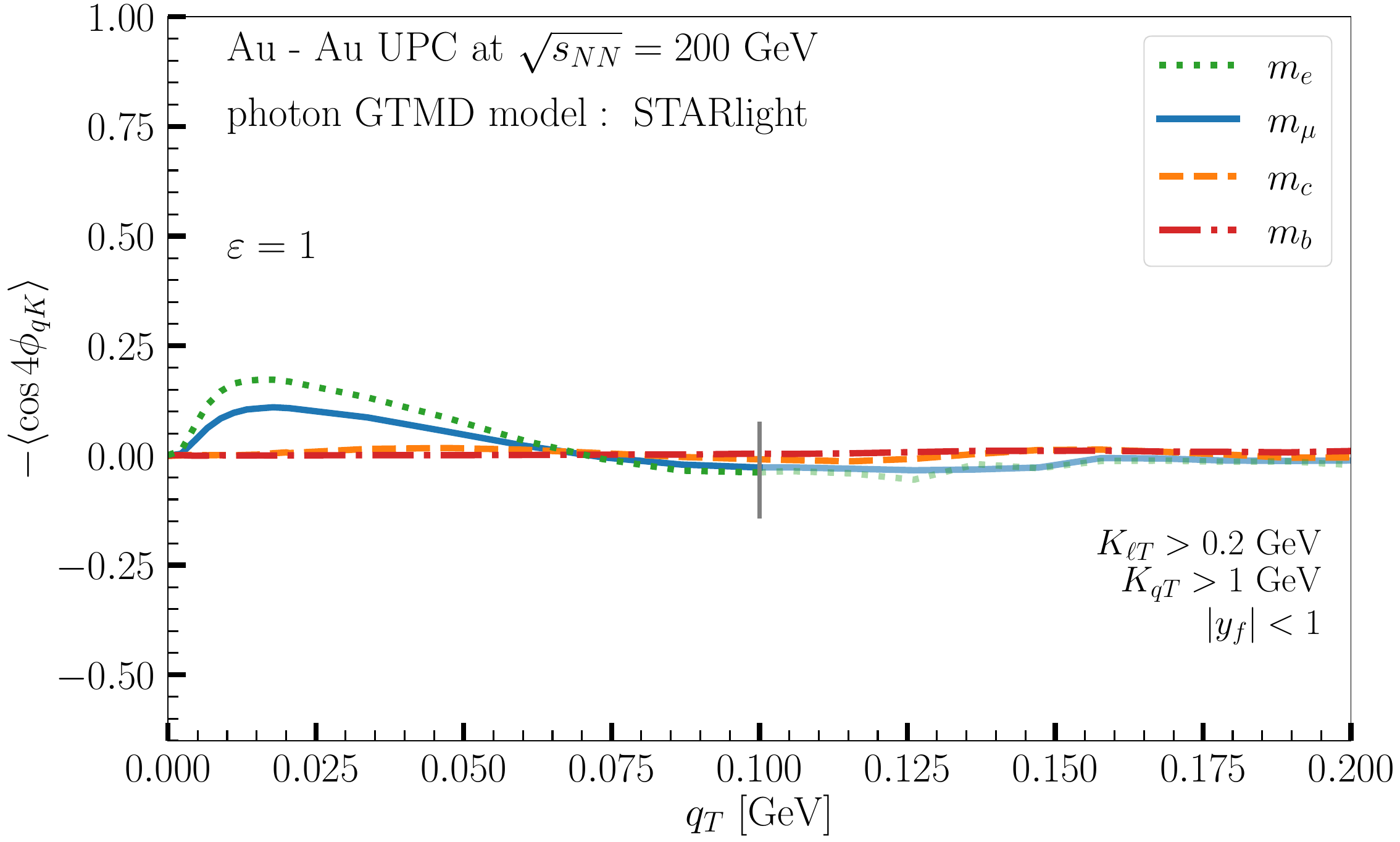}}\hspace{1cm}
\subfloat[\label{fig: mass 2phiqK}]
{\includegraphics[width=.333\linewidth, keepaspectratio]{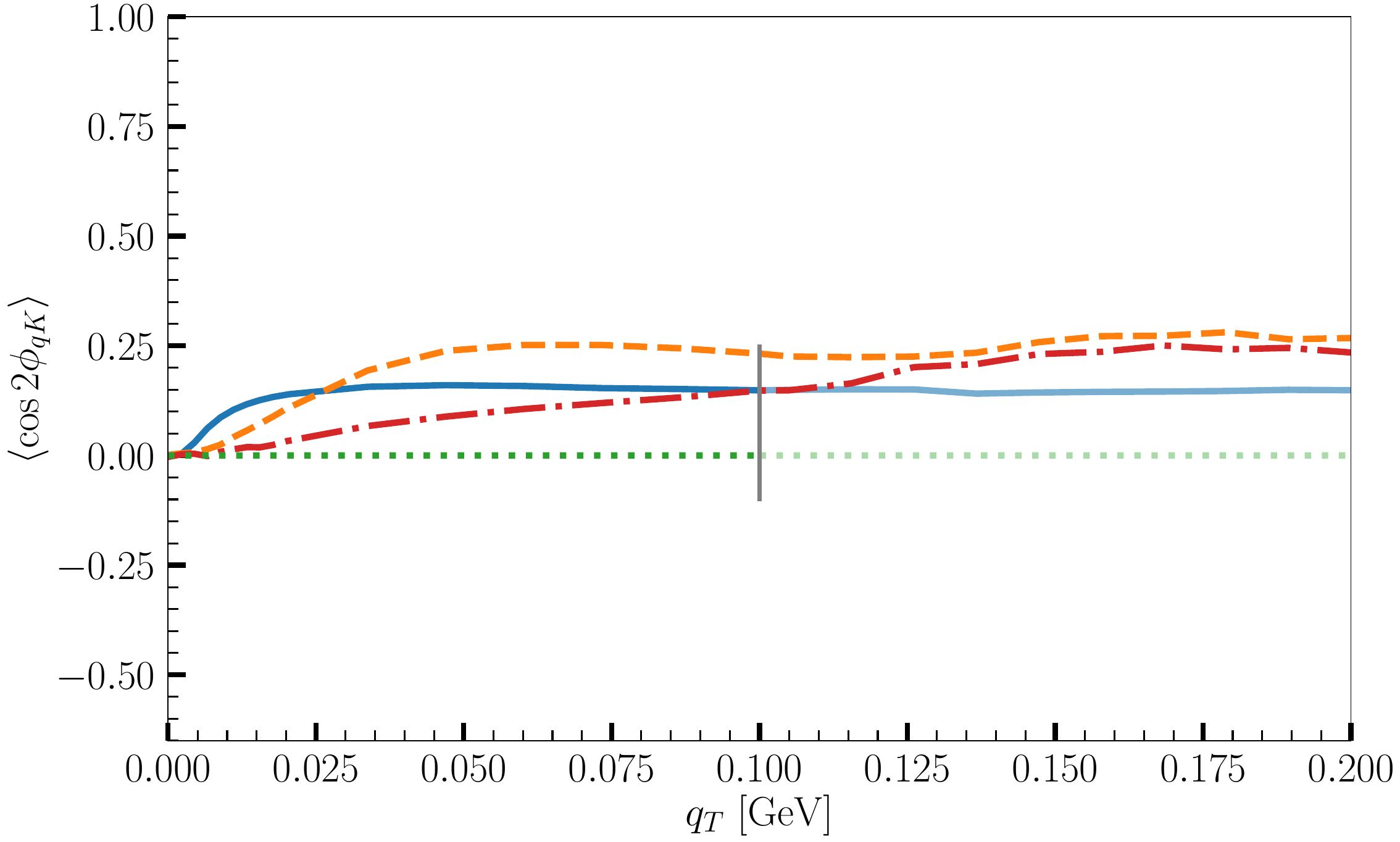}}\\
\subfloat[\label{fig: mass 2phiqD}]
{\includegraphics[width=.333\linewidth, keepaspectratio]{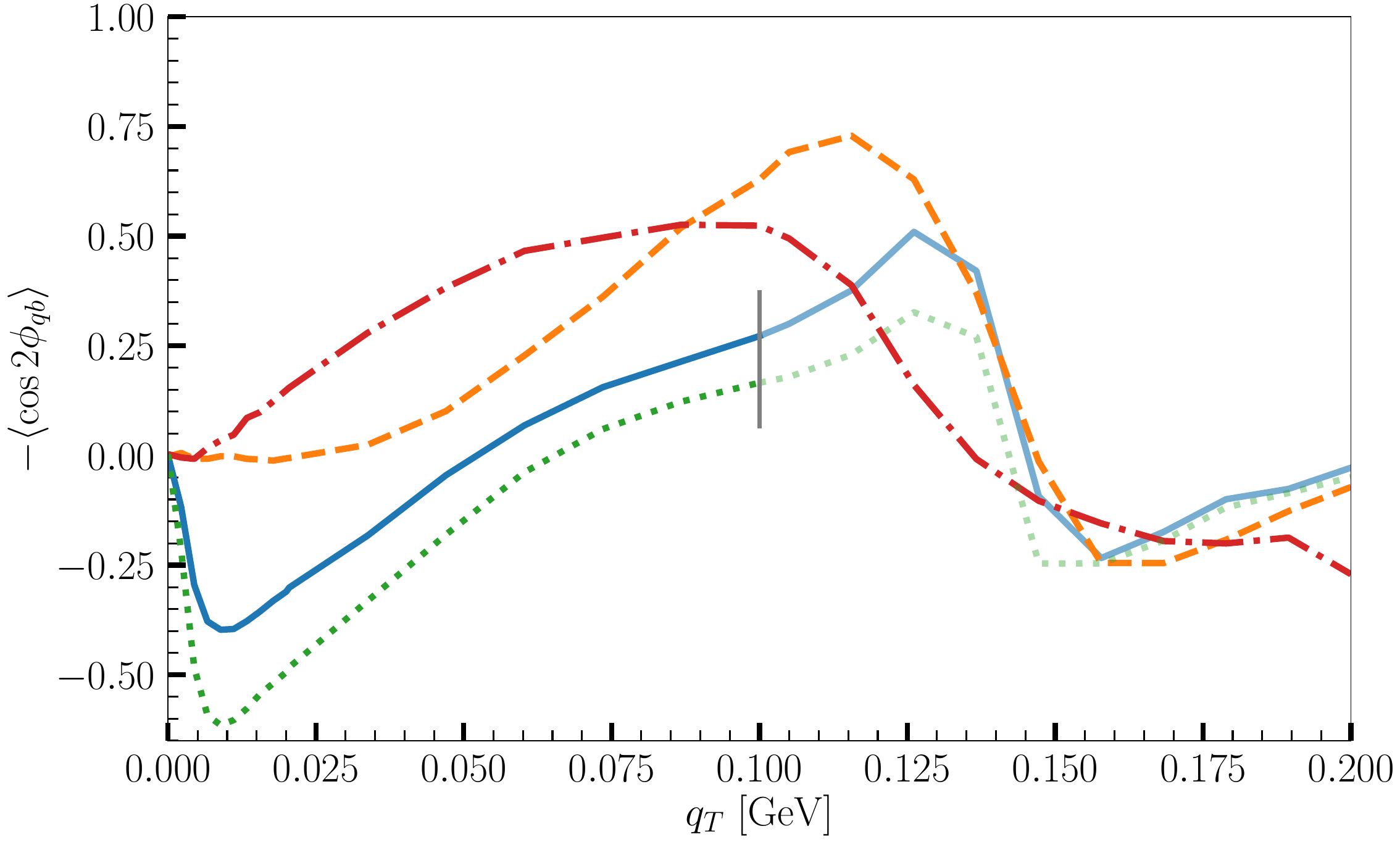}}\hfill
\subfloat[]
{\includegraphics[width=.333\linewidth, keepaspectratio]{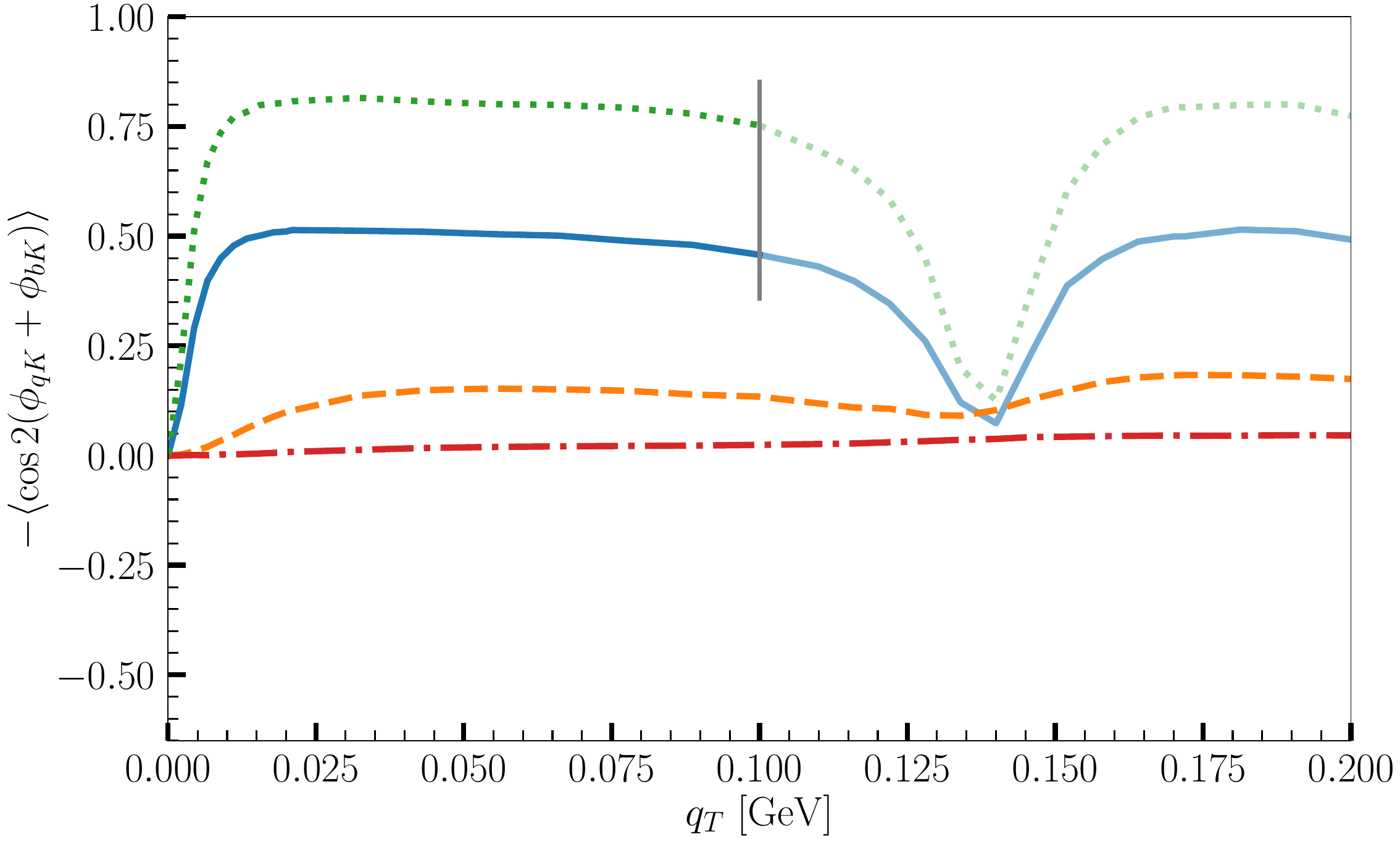}}\hfill
\subfloat[\label{fig: mass 2phiDK}]
{\includegraphics[width=.333\linewidth, keepaspectratio]{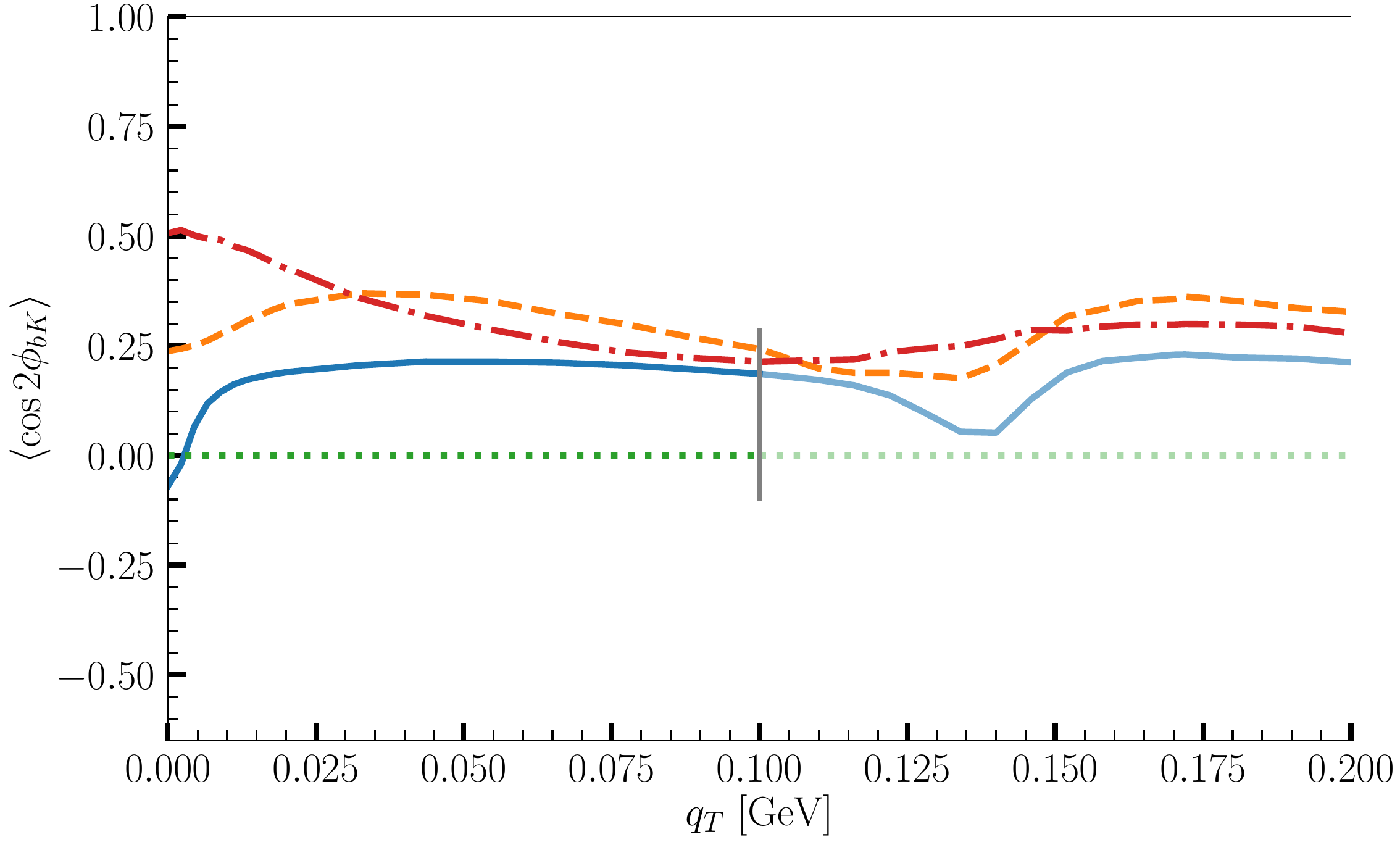}}\\
\subfloat[\label{fig: mass 4phiqD}]
{\includegraphics[width=.333\linewidth, keepaspectratio]{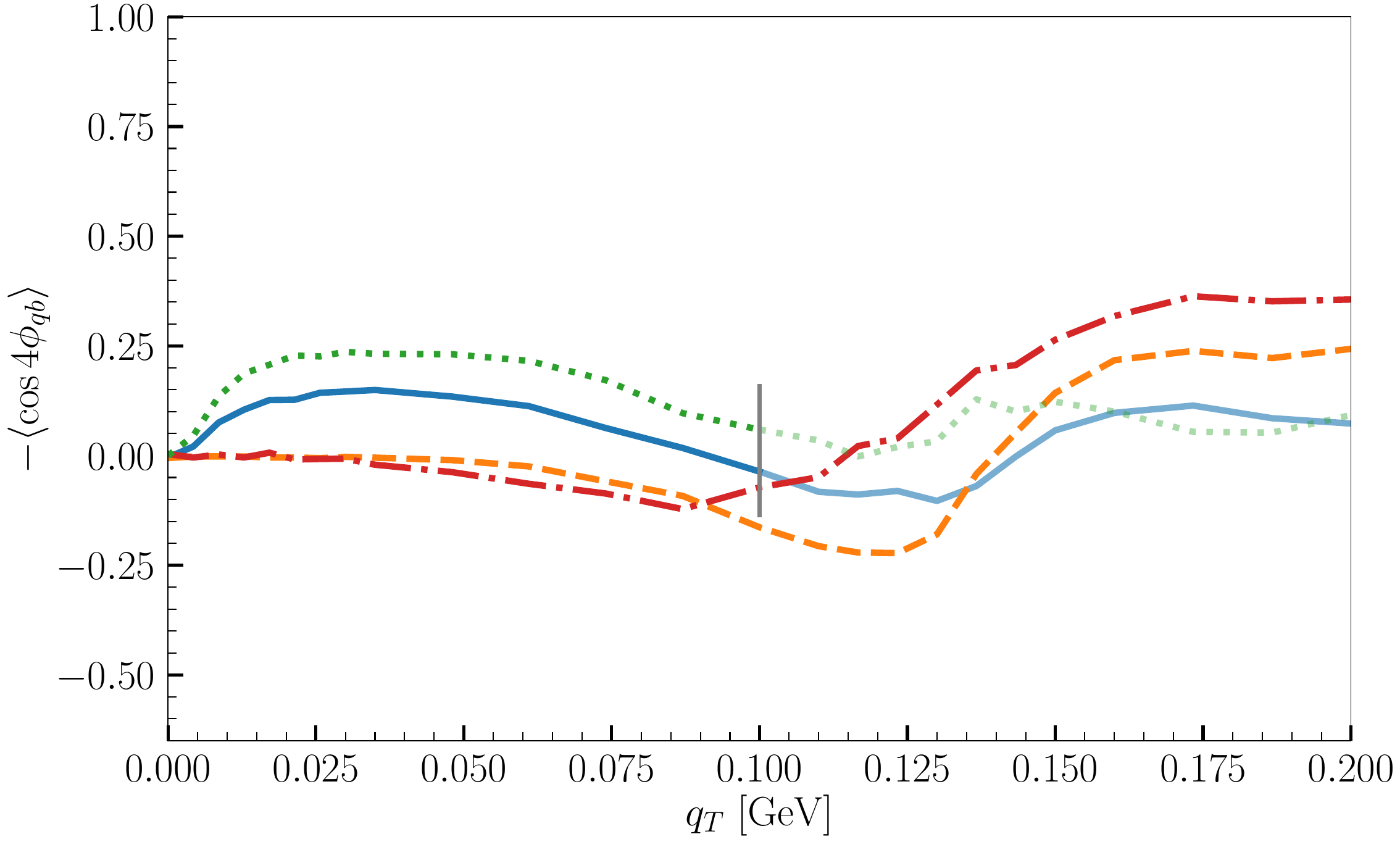}}\hspace{1cm}
\subfloat[]
{\includegraphics[width=.333\linewidth, keepaspectratio]{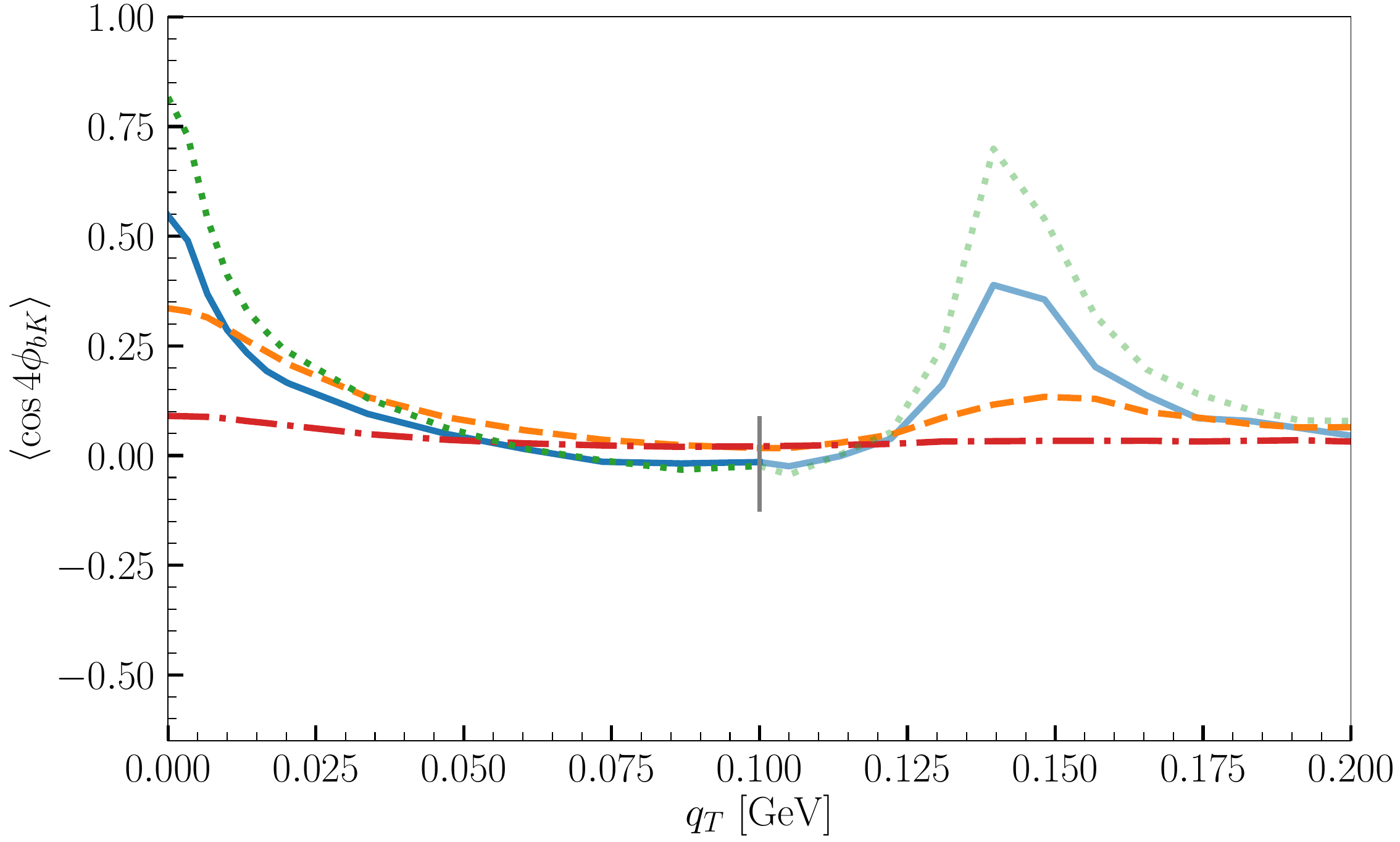}}\hfill
\caption{\it Mass dependence of the various asymmetries using the STARlight FF. For all panels, the green dotted line corresponds to $e^+ e^-$ production, blue solid to $\mu^+ \mu^-$, orange dashed to $c \bar c$, and the red dash-dotted one to $b \bar b$. The vertical lines correspond to the points after which we cannot trust the factorization for the lighter states.} 
\label{fig: asymmetries masses}
\end{center}
\end{figure}

In Fig.~\ref{fig: asymmetries ee} we present for $e^+ e^-$ production at RHIC kinematics all the dominant asymmetries of Eq.~\eqref{eq: cs dileqepton b space} together with $\langle \cos4\phi_{qb}\rangle$, which can be sizable upon inclusion of feed-in contributions. 
The $\langle \cos 4\phi_{qK} \rangle$ is the only asymmetry that survives in the forward limit. It behaves like $q_\sT^4$ at small $q_\sT$, while at high $q_\sT$ off-forward contributions significantly suppress this asymmetry compared to its forward limit, with
the suppression enhanced by the feed-in.
All other asymmetries (including those not presented here) involve the azimuthal angle $\phi_b$. The $\langle \cos2\phi_{qb}\rangle$ and $\langle \cos2(\phi_{qK} + \phi_{bK})\rangle$ (dashed lines) are obtained using Eq.~\eqref{eq: UPC b-integrated formula - 2phi}, while $\langle \cos4\phi_{bK}\rangle$ and $\langle \cos4\phi_{qb}\rangle$ (dash-dotted lines) are derived using Eq.~\eqref{eq: UPC b-integrated formula - 4phi}.
Among these, $\langle \cos4\phi_{bK}\rangle$ is the only asymmetry that is nonzero at $q_\sT = 0$ since the angle $\phi_{bK}$ is still definable for that $q_\sT$ value. In contrast, all other asymmetries go to zero as $q_\sT^n$, with $n$ being the number of $\phi_q$ in the asymmetry (or equivalently the order of the harmonic).
For larger values of $q_\sT$ the most sizable asymmetry is $\langle \cos2(\phi_{qK} + \phi_{bK}) \rangle$.
Despite the absence of a corresponding term in $\Delta_\sT$ space, the $\langle \cos 4 \phi_{qb} \rangle$ asymmetry turns out to be sizable (up to $25\%$) solely from feed-in contributions.
A final observation on Fig.~\ref{fig: asymmetries ee} concerns the comparison with the findings of~\cite{Shi:2024gex}. 
Although the authors provide similar figures to Fig.~\ref{fig: asymmetries ee}, they only considered fixed $b_\sT$, while our results are obtained by integrating separately the numerator and denominator over the impact parameter for $b_\sT \geq b_{\min}=2R_A$.
Hence, a direct comparison between the figures should not be made. 
We have however evaluated at fixed $b_\sT$ the same asymmetries discussed here, and find full agreement with the results of~\cite{Shi:2024gex}.

In Fig.~\ref{fig: asymmetries masses} we present the mass dependence of the asymmetries. For clarity, we show each asymmetry in a separate panel. Moreover, we changed the sign of $\langle \cos2\phi_{qb}\rangle$ compared to the previous figure (as also indicated in the $y$-labels) to ensure that all results are predominantly positive. As before, for the kinematics considered which are the same as for Fig.~\ref{fig: mass dependence}, we trust the factorization up to $q_\sT = 0.1~{\rm GeV}$ for the production of the light particles, whereas the $q_\sT$ range may be broader for the heavy quarks. 
However, we have verified that all asymmetries respect the physical bounds also beyond the $q_\sT$ ranges shown in the figures.
In addition to those discussed in Fig.~\ref{fig: asymmetries ee} for the $e^+ e^-$ production, we have two asymmetries that are significantly nonzero when the produced particles are massive: $\langle \cos2\phi_{qK} \rangle$ and $\langle \cos2\phi_{b K} \rangle$, which are respectively given in Figs.~\ref{fig: mass 2phiqK} and~\ref{fig: mass 2phiDK}. 
We have found that already the muon mass, at least at the energies considered here, is sufficient to make these asymmetries observable.
In general, we can see that the expectations of all asymmetries between the two dilepton systems presented in this work, although similar, differ more than those of the normalized cross sections, motivating experimental studies that investigate such differences specifically.
Besides, the heavier states, given by the $c \bar c$ and $b \bar b$ systems, lead to remarkably distinct results from those of the dilepton ones. Among them, we have that $\langle \cos4\phi_{qK} \rangle$ is strongly suppressed in open heavy-quark production, with the dominant $\phi_{qK}$ dependence thus being the $\cos2\phi_{qK}$. Note that for muons both terms are rather sizable.
Furthermore, $\langle \cos2(\phi_{qK} + \phi_{bK})\rangle$ is negligible for heavy quarks, while it is significantly nonzero in the dilepton case.
It is important to highlight that the asymmetries in open-heavy quark production present other general features that might be observed in experiments. In particular, we have that $\langle \cos2\phi_{qb}\rangle$ and $\langle \cos4\phi_{qb}\rangle$ (the latter only driven by feed-in contributions as already mentioned) have opposite signs between dilepton and heavy-quark productions for $q_\sT<0.1~{\rm GeV}$, while $\langle \cos2\phi_{qK}\rangle$ is positive for heavy-quark production, and comparable (except for $q_\sT = 0$) to the $\mu^+ \mu^-$ one.
Hence, even if the measurement of heavy quarks is less clean, we believe that the results shown here are worth experimental investigation.

As a final comment on the various asymmetries, we address the bumps and dips that some of the curves display around $q_\sT \sim 0.125~\textrm{-}~0.15$ that have the same origin as the oscillation pointed out in Fig.~\ref{fig: mass dependence} for the cross section. 
The effects grow or shrink depending on which value is chosen for $b_{\min}$ (see Eqs.~\eqref{eq: UPC b-integrated formula - 0}~-~\eqref{eq: UPC b-integrated formula - 4phi}).
Their positions also shift somewhat with $b_{\min}$, but they remain located roughly around the same values of $q_\sT$.
While for the cross section the oscillation is significant mostly in the (light) dilepton production, for the asymmetries the same effect is enhanced enough to be observable in the (heavy) quark-pair production as well. 
For instance, $\langle \cos2\phi_{qb}\rangle$ and $\langle \cos4\phi_{qb}\rangle$ show a sign change within the $q_\sT$ range specified above which can be tested in experiments, e.g.~at LHC.

\begin{figure}[t]
\begin{center}
\subfloat[\label{fig: dcs LHC}]
{\includegraphics[width=.495\linewidth, keepaspectratio]{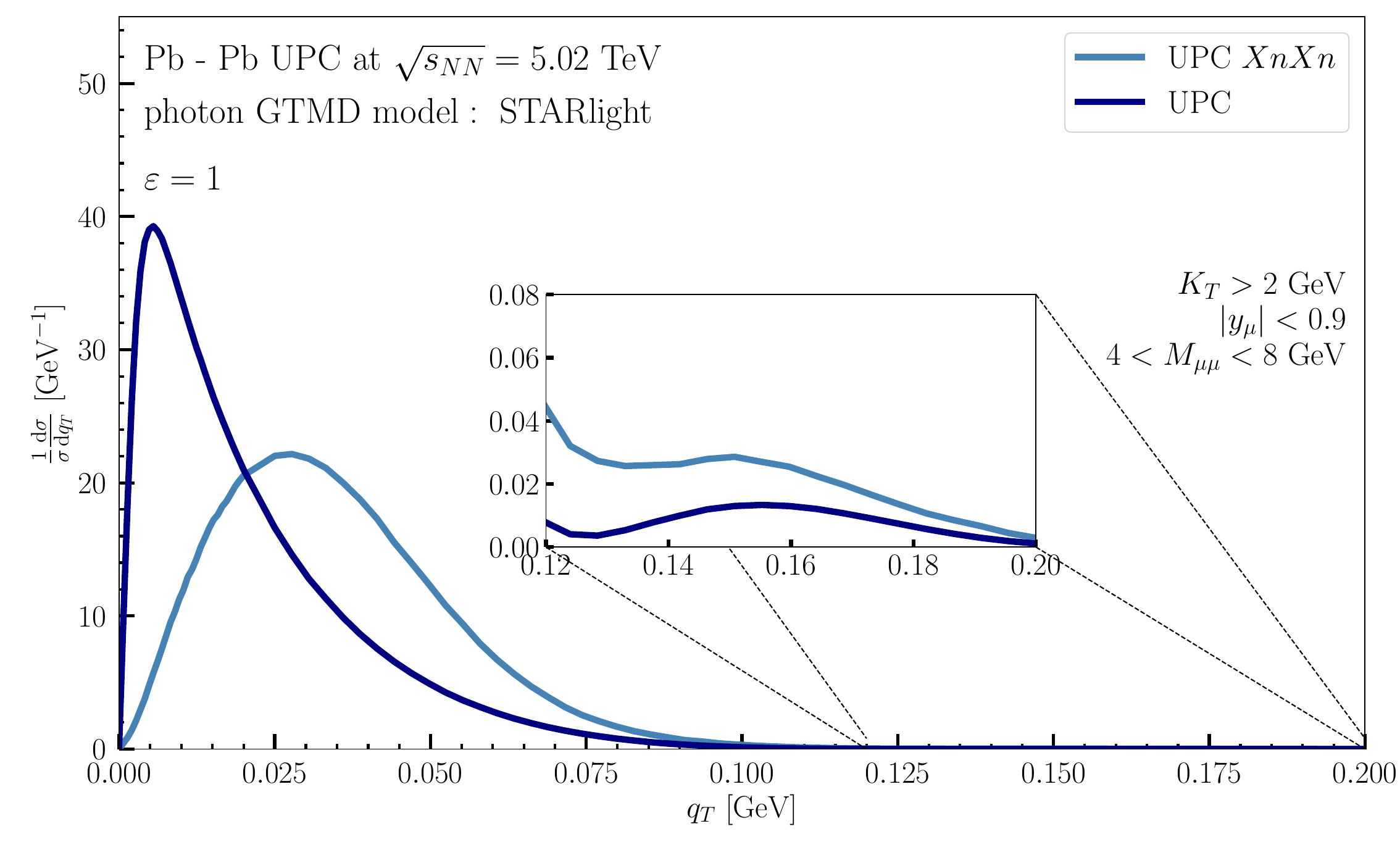}}\hfill
\subfloat[\label{fig: asymmetries LHC}]
{\includegraphics[width=.495\linewidth, keepaspectratio]{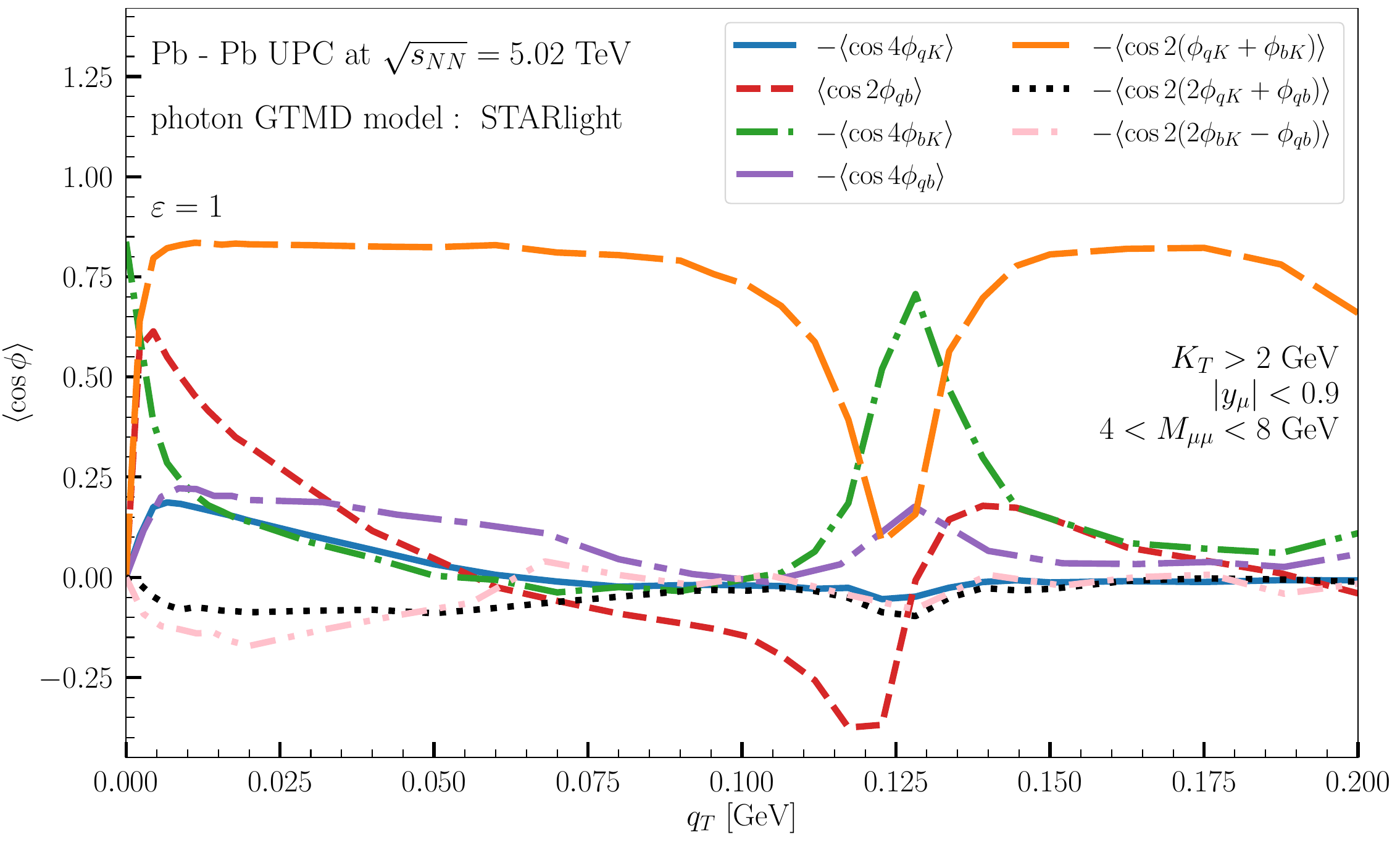}}
\caption{\it (a) Normalized differential cross section as a function of $q_\sT$ for $\mu^+ \mu^-$ production in Pb~-~Pb UPCs at $\sqrt{s_{NN}} = 5.02~{\rm TeV}$.  
(b) Several asymmetries for $\mu^+ \mu^-$ production in Pb~-~Pb UPCs at LHC. Kinematical constraints (same for both panels) are given in the legend.}
\label{fig: LHC predictions}
\end{center}
\end{figure}

In Fig.~\ref{fig: LHC predictions} we present the normalized cross section and various asymmetries relevant for LHC kinematics. In particular, these predictions are obtained for dimuon production at mid rapidity, with an invariant mass of the dilepton system of $4 < M_{\mu\mu} < 8~{\rm GeV}$, which excludes quarkonium resonances. The general behavior of the dimuon production at mid-rapidity at LHC is analogous to that of dielectron production at RHIC discussed above.
In particular, we have that the bulk of the cross section, Fig.~\ref{fig: dcs LHC}, for both full UPC (all neutron emissions channels included) and $XnXn$ (one or more neutrons emitted) is found at $q_\sT < 100~{\rm MeV}$, with the first oscillation expected at $q_\sT \approx 150~{\rm MeV}$. We point out that the principal peak of the distribution is also found at $q_\sT < 100~{\rm MeV}$ for other kinematical selections, e.g.~in a forward detector.
Regarding the asymmetries, in Fig.~\ref{fig: asymmetries LHC} we have included all the asymmetries that can be generated by the feed-in mechanism at the first order and are greater than $10\%$. Due to the higher $K_\sT$ selection, mass effects in dimuon production are negligible, which causes the suppression of asymmetries of the $\langle \cos2{\phi_{aK}} \rangle$ type at the LHC. Again, we observe the presence of bumps and dips in the asymmetries around $q_\sT \sim 125~\textrm{-}~150\ {\rm MeV}$.
However, in contrast with the cross section where the rapidity region does not significantly affect our predictions, for the asymmetries we have found that the mid-rapidity region is a better choice to measure the asymmetries. Indeed, in the forward rapidity region most of these asymmetries become negligible, with solely $\langle \cos2(\phi_{qK} + \phi_{bK}) \rangle$, $\langle \cos4\phi_{bK} \rangle$ and $\langle \cos2\phi_{bq} \rangle$ remaining above $10\%$.

\section{Conclusions} 
\label{sec conclusions}

In this paper we have investigated the differential cross sections for lepton and heavy-quark pair production in UPCs, as well as their 
azimuthal modulations arising from the collisions of unpolarized and polarized photons emitted by 
highly charged ions. This has been the subject of many earlier investigations, but thus far only Ref.~\cite{Shi:2024gex} considered the azimuthal modulations in the cross section differential in both the sum and difference of the transverse momenta of the final state particle pair, i.e.\ in $\bm K_\sT$ and $\bm q_\sT$, as well as in the transverse impact parameter $\bm b_\sT$. Wherever we could compare we find agreement between our results for the azimuthal modulations in the relative angles between these three vectors and those given in Ref.~\cite{Shi:2024gex} and in other papers on this topic~\cite{Vidovic:1992ik,Hencken:1994my,Zha:2018ywo,Li:2019sin,Xiao:2020ddm,Wang:2021kxm,Mazurek:2021ahz,Shao:2022stc,Shao:2023zge}, differing only in mass terms and feed-in contributions (and one sign that we discuss in Appendix \ref{sec: cross check}). More specifically, compared to these other works we have systematically included mass terms that are of relevance in the production of heavier leptons and quarks, we have separated effects that arise from the anisotropies of the GTMDs from those that do not, and rather than presenting asymmetries at a fixed impact parameter $b_\sT$ (which can not be determined exactly anyway) we have chosen to show asymmetries integrated over all $b_\sT \geq b_{\min} = 2R_A$. We have pointed out that if one includes the terms that arise from the $k_\sT \cdot \Delta_\sT$ dependence of the GTMDs (or equivalently the $q_\sT \cdot \Delta_\sT$ dependence of the convolutions of two GTMDs), one loses the one-to-one relation between the angular modulations in $\phi_\Delta$ and $\phi_b$, leading to a feed-in mechanism among harmonics of different order, complicating model predictions and generating ever higher harmonics. In this way one finds that despite the absence of a corresponding term in $\Delta_\sT$ space, the $\langle \cos 4 \phi_{qb} \rangle$ asymmetry turns out to be sizable (in the model study up to $25\%$) solely from feed-in contributions. We have also found that although the cross section differential in the off-forwardness contains sine modulations, the Fourier transformed cross section in impact parameter space does not (ignoring possible effects from gauge links). The former terms do need to be included, however, as they feed into cosine modulations and without them positivity of some of these asymmetries is found to be violated. We emphasize that the off-forwardness $\Delta_\sT$ which on average will be small for UPCs is not directly observable in the UPC processes, unless the recoil momenta of the nuclei are measured precisely, hence, the complications from the Fourier transform to impact parameter space are unavoidable. 

We have included terms proportional to the mass of the produced particles because they are relevant for muon, charm and bottom quark production. We have shown that the normalized differential cross section changes considerably with the produced particle mass when comparing the open-quark to dilepton productions, whereas the difference between muons and electrons is small enough that one might increase statistics by combining the two data. The same does not apply to the angular modulations, where we can observe significant differences for all the final states considered in this work. In particular, we point out that the $\cos2\phi_{qK}$ asymmetry is the dominant $\phi_{qK}$ dependence for charm and bottom and not the $\cos4\phi_{qK}$ observed for electrons. Furthermore, we observe that the $\cos2\phi_{qb}$ and $\cos4\phi_{qb}$ asymmetries change sign if one goes from light-leptons to heavy-quarks production, while $\cos2(\phi_{qK} + \phi_{bK})$ is suppressed.
Such mass effects should be discernible in the UPC data because the differences are expected to be larger than the experimental errors (at least those of the electron data of RHIC). For the numerical study we have considered several models for the photon GTMD correlator, but they all lead to comparable results for UPCs with impact parameters for which the two nuclei do not overlap. For peripheral and more central collisions these results will not be applicable and different models may lead to very distinct results.
Finally, since the models all lead to anisotropic terms in the GTMDs, which one can reduce by hand, we found that RHIC data slightly favor models that do not suppress feed-in contributions. Hopefully all these observations can be explored with UPC data from LHC as well.

\section{Acknowledgments}
This work is supported by the European Union ``Next Generation EU'' program through the Italian PRIN 2022 grant n.~20225ZHA7W.
We thank the BNL EIC Theory Institute for the hospitality during a two-week stay when part of this work was developed.

\appendix

\section{Comparison to cross section expressions in the literature}
\label{sec: cross check}

In this appendix, we compare our expressions given in Eq.~\eqref{eq: structure functions explicit - 0phiK} for $M_\ell \ll M_\sT$ with the equivalent ones in Refs.~\cite{Vidovic:1992ik, Klein:2020jom, Li:2019sin}.
We first consider the formula in Eq.~(31) of Ref.~\cite{Vidovic:1992ik}, which depends on
\begin{align}
    {\cal A}^{\rm VGBS} & = 
        (\bm \kappa_{1\sT} \cdot \bm \kappa_{2\sT})\, (\bm \kappa^\prime_{1\sT} \cdot \bm \kappa^\prime_{2\sT})\,\sigma_{\myparallel} + (\bm \kappa_{1\sT}\times \bm \kappa_{2\sT}) \cdot (\bm \kappa^\prime_{1\sT} \times \bm \kappa^\prime_{2\sT})\, \sigma_{\perp} \nonumber\\
    & = 
        (\bm \kappa_{1\sT} \cdot \bm \kappa_{2\sT})\, (\bm \kappa^\prime_{1\sT} \cdot \bm \kappa^\prime_{2\sT})\,\sigma_{\myparallel} + \left[ (\bm \kappa_{1\sT} \cdot \bm \kappa^\prime_{1\sT})\, (\bm \kappa_{2\sT} \cdot \bm \kappa^\prime_{2\sT}) - (\bm \kappa_{1\sT} \cdot \bm \kappa^\prime_{2\sT})\, (\bm \kappa_{2\sT} \cdot \bm \kappa^\prime_{1\sT}) \right] \sigma_{\perp}\, ,
\end{align}
with $\sigma_\myparallel$ and $\sigma_\perp$ being the scalar and pseudoscalar components of the unpolarized cross section for the process $\gamma\gamma\to \ell^+\ell^-$, in which the polarizations of the two initial photons are taken to be parallel and perpendicular to each other, respectively. 
If one neglects the final lepton masses, we have that ${\sigma_\myparallel = \sigma_\perp \equiv \sigma_0/2}$.
Thus, by employing Eq.~\eqref{eq: kappa definition}, we have that
\begin{align}
    2\sigma_0^{-1}\, {\cal A}^{\rm VGBS} & = 
    \big (\bm \kappa_{1\sT} \cdot \bm \kappa^\prime_{1\sT} \big)\, \big(\bm \kappa_{2\sT} \cdot \bm \kappa^\prime_{2\sT} \big) + 
    \left[\left( \bm k_{1\sT} \cdot \bm k_{2\sT} - \frac{\bm \Delta_\sT^2}{4} \right)^2 - \frac{1}{4} \big( \bm \Delta_\sT \cdot k_{1\sT} - \bm \Delta_\sT \cdot \bm k_{2\sT} \big)^2 \right] \nonumber\\
    & \phantom{=} - \left[ \left( \bm k_{1\sT} \cdot \bm k_{2\sT} + \frac{\bm \Delta_\sT^2}{4} \right)^2 - \frac{1}{4} \big( \bm \Delta_\sT \cdot k_{1\sT} + \bm \Delta_\sT \cdot \bm k_{2\sT} \big)^2 \right] \nonumber\\
    & = \big(\bm \kappa_{1\sT} \cdot \bm \kappa^\prime_{1\sT} \big)\, \big(\bm \kappa_{2\sT} \cdot \bm \kappa^\prime_{2\sT} \big) - \frac{\bm \Delta_\sT^2\, |\bm k_{1\sT}|\, |\bm k_{2\sT}|}{2} \big( \cos\phi_{12} - \cos(\phi_{\Delta1} + \phi_{\Delta2}) \big) \, .
\label{eq: Vidovic amplitude}
\end{align}
On the other hand, from Eq.~\eqref{eq: cs dileqepton DeltaT-space} and employing Eq.~\eqref{eq: GTMDs EPA}, we have that
\begin{align}
   2 \sigma_0^{-1} {\cal A} = \big(\bm \kappa_{1\sT} \cdot \bm \kappa^\prime_{1\sT} \big)\, \big(\bm \kappa_{2\sT} \cdot \bm \kappa^\prime_{2\sT} \big) + \frac{\bm \Delta_\sT^2\, |\bm k_{1\sT}|\, |\bm k_{2\sT}|}{2} \big( \cos\phi_{12} - \cos(\phi_{\Delta1} + \phi_{\Delta2}) \big) \, ,
\label{eq: our amplitude}
\end{align}
with $\sigma_0 = 2 \frac{\alpha^2}{s_{NN} M_\sT^2} [z^2 + (1 - z)^2]$. The first term comes from the convolution ${\cal C}[{\cal F}^\gamma_1 {\cal F}^\gamma_1]$, whereas the other terms are related to the convolutions involving ${\cal F}^\gamma_4$.
We observe that there is a discrepancy between the result in Eq.~\eqref{eq: our amplitude} and that of Ref.~\cite{Vidovic:1992ik}. The difference is not related to differences in the definition of $\Delta_\sT$ nor ${\cal F}_4^\gamma$ since both appear squared.
Nevertheless, Eq.~\eqref{eq: our amplitude} agrees with the more recent works, namely Eq.~(3) of~\cite{Li:2019sin} and Eq.~(26) of~\cite{Klein:2020jom}.

\bibliography{bibliography}
\end{document}